\documentclass[%
reprint,
superscriptaddress,
showpacs,
 amsmath,amssymb,
 aps,
 prl,
]{revtex4-1}

\usepackage{graphicx}
\usepackage{dcolumn}
\usepackage{bm}
\usepackage{amsmath}
\usepackage{autobreak}
\usepackage{hyperref}
\usepackage[mathlines]{lineno}
\usepackage{natbib}
\usepackage{multirow}

\def\chie{$\chi$-$e$ }
\def\mchi{$m_{\chi}$ }

\begin{document}

\preprint{APS/123-QED}

\title{Experimental Limits on Solar Reflected Dark Matter with a New Approach on Accelerated-Dark-Matter--Electron Analysis in Semiconductors}

\author{Z.~Y.~Zhang}
\affiliation{Key Laboratory of Particle and Radiation Imaging (Ministry of Education) and Department of Engineering Physics, Tsinghua University, Beijing 100084}
\author{L.~T.~Yang}
\email{Corresponding author: yanglt@mail.tsinghua.edu.cn}
\affiliation{Key Laboratory of Particle and Radiation Imaging (Ministry of Education) and Department of Engineering Physics, Tsinghua University, Beijing 100084}
\author{Q.~Yue}
\email{Corresponding author: yueq@mail.tsinghua.edu.cn}
\affiliation{Key Laboratory of Particle and Radiation Imaging (Ministry of Education) and Department of Engineering Physics, Tsinghua University, Beijing 100084}
\author{K.~J.~Kang}
\affiliation{Key Laboratory of Particle and Radiation Imaging (Ministry of Education) and Department of Engineering Physics, Tsinghua University, Beijing 100084}
\author{Y.~J.~Li}
\affiliation{Key Laboratory of Particle and Radiation Imaging (Ministry of Education) and Department of Engineering Physics, Tsinghua University, Beijing 100084}

\author{H.~P.~An}
\affiliation{Key Laboratory of Particle and Radiation Imaging (Ministry of Education) and Department of Engineering Physics, Tsinghua University, Beijing 100084}
\affiliation{Department of Physics, Tsinghua University, Beijing 100084}

\author{Greeshma~C.}
\altaffiliation{Participating as a member of TEXONO Collaboration}
\affiliation{Institute of Physics, Academia Sinica, Taipei 11529}

\author{J.~P.~Chang}
\affiliation{NUCTECH Company, Beijing 100084}

\author{Y.~H.~Chen}
\affiliation{YaLong River Hydropower Development Company, Chengdu 610051}
\author{J.~P.~Cheng}
\affiliation{Key Laboratory of Particle and Radiation Imaging (Ministry of Education) and Department of Engineering Physics, Tsinghua University, Beijing 100084}
\affiliation{College of Nuclear Science and Technology, Beijing Normal University, Beijing 100875}
\author{W.~H.~Dai}
\affiliation{Key Laboratory of Particle and Radiation Imaging (Ministry of Education) and Department of Engineering Physics, Tsinghua University, Beijing 100084}
\author{Z.~Deng}
\affiliation{Key Laboratory of Particle and Radiation Imaging (Ministry of Education) and Department of Engineering Physics, Tsinghua University, Beijing 100084}
\author{C.~H.~Fang}
\affiliation{College of Physics, Sichuan University, Chengdu 610065}
\author{X.~P.~Geng}
\affiliation{Key Laboratory of Particle and Radiation Imaging (Ministry of Education) and Department of Engineering Physics, Tsinghua University, Beijing 100084}
\author{H.~Gong}
\affiliation{Key Laboratory of Particle and Radiation Imaging (Ministry of Education) and Department of Engineering Physics, Tsinghua University, Beijing 100084}
\author{Q.~J.~Guo}
\affiliation{School of Physics, Peking University, Beijing 100871}
\author{T.~Guo}
\affiliation{Key Laboratory of Particle and Radiation Imaging (Ministry of Education) and Department of Engineering Physics, Tsinghua University, Beijing 100084}
\author{X.~Y.~Guo}
\affiliation{YaLong River Hydropower Development Company, Chengdu 610051}
\author{L.~He}
\affiliation{NUCTECH Company, Beijing 100084}
\author{S.~M.~He}
\affiliation{YaLong River Hydropower Development Company, Chengdu 610051}
\author{J.~W.~Hu}
\affiliation{Key Laboratory of Particle and Radiation Imaging (Ministry of Education) and Department of Engineering Physics, Tsinghua University, Beijing 100084}
\author{H.~X.~Huang}
\affiliation{Department of Nuclear Physics, China Institute of Atomic Energy, Beijing 102413}
\author{T.~C.~Huang}
\affiliation{Sino-French Institute of Nuclear and Technology, Sun Yat-sen University, Zhuhai 519082}
\author{L.~Jiang}
\affiliation{Key Laboratory of Particle and Radiation Imaging (Ministry of Education) and Department of Engineering Physics, Tsinghua University, Beijing 100084}
\author{S.~Karmakar}
\altaffiliation{Participating as a member of TEXONO Collaboration}
\affiliation{Institute of Physics, Academia Sinica, Taipei 11529}

\author{H.~B.~Li}
\altaffiliation{Participating as a member of TEXONO Collaboration}
\affiliation{Institute of Physics, Academia Sinica, Taipei 11529}
\author{H.~Y.~Li}
\affiliation{College of Physics, Sichuan University, Chengdu 610065}
\author{J.~M.~Li}
\affiliation{Key Laboratory of Particle and Radiation Imaging (Ministry of Education) and Department of Engineering Physics, Tsinghua University, Beijing 100084}
\author{J.~Li}
\affiliation{Key Laboratory of Particle and Radiation Imaging (Ministry of Education) and Department of Engineering Physics, Tsinghua University, Beijing 100084}
\author{Q.~Y.~Li}
\affiliation{College of Physics, Sichuan University, Chengdu 610065}
\author{R.~M.~J.~Li}
\affiliation{College of Physics, Sichuan University, Chengdu 610065}
\author{X.~Q.~Li}
\affiliation{School of Physics, Nankai University, Tianjin 300071}
\author{Y.~L.~Li}
\affiliation{Key Laboratory of Particle and Radiation Imaging (Ministry of Education) and Department of Engineering Physics, Tsinghua University, Beijing 100084}
\author{Y.~F.~Liang}
\affiliation{Key Laboratory of Particle and Radiation Imaging (Ministry of Education) and Department of Engineering Physics, Tsinghua University, Beijing 100084}
\author{B.~Liao}
\affiliation{College of Nuclear Science and Technology, Beijing Normal University, Beijing 100875}
\author{F.~K.~Lin}
\altaffiliation{Participating as a member of TEXONO Collaboration}
\affiliation{Institute of Physics, Academia Sinica, Taipei 11529}
\author{S.~T.~Lin}
\affiliation{College of Physics, Sichuan University, Chengdu 610065}
\author{J.~X.~Liu}
\affiliation{Key Laboratory of Particle and Radiation Imaging (Ministry of Education) and Department of Engineering Physics, Tsinghua University, Beijing 100084}
\author{S.~K.~Liu}
\affiliation{College of Physics, Sichuan University, Chengdu 610065}
\author{Y.~D.~Liu}
\affiliation{College of Nuclear Science and Technology, Beijing Normal University, Beijing 100875}
\author{Y.~Liu}
\affiliation{College of Physics, Sichuan University, Chengdu 610065}
\author{Y.~Y.~Liu}
\affiliation{College of Nuclear Science and Technology, Beijing Normal University, Beijing 100875}
\author{H.~Ma}
\affiliation{Key Laboratory of Particle and Radiation Imaging (Ministry of Education) and Department of Engineering Physics, Tsinghua University, Beijing 100084}
\author{Y.~C.~Mao}
\affiliation{School of Physics, Peking University, Beijing 100871}
\author{Q.~Y.~Nie}
\affiliation{Key Laboratory of Particle and Radiation Imaging (Ministry of Education) and Department of Engineering Physics, Tsinghua University, Beijing 100084}
\author{J.~H.~Ning}
\affiliation{YaLong River Hydropower Development Company, Chengdu 610051}
\author{H.~Pan}
\affiliation{NUCTECH Company, Beijing 100084}
\author{N.~C.~Qi}
\affiliation{YaLong River Hydropower Development Company, Chengdu 610051}
\author{J.~Ren}
\affiliation{Department of Nuclear Physics, China Institute of Atomic Energy, Beijing 102413}
\author{X.~C.~Ruan}
\affiliation{Department of Nuclear Physics, China Institute of Atomic Energy, Beijing 102413}
\author{M.~K.~Singh}
\altaffiliation{Participating as a member of TEXONO Collaboration}
\affiliation{Institute of Physics, Academia Sinica, Taipei 11529}
\affiliation{Department of Physics, Banaras Hindu University, Varanasi 221005}
\author{T.~X.~Sun}
\affiliation{College of Nuclear Science and Technology, Beijing Normal University, Beijing 100875}
\author{C.~J.~Tang}
\affiliation{College of Physics, Sichuan University, Chengdu 610065}
\author{Y.~Tian}
\affiliation{Key Laboratory of Particle and Radiation Imaging (Ministry of Education) and Department of Engineering Physics, Tsinghua University, Beijing 100084}
\author{G.~F.~Wang}
\affiliation{College of Nuclear Science and Technology, Beijing Normal University, Beijing 100875}
\author{J.~Z.~Wang}
\affiliation{Key Laboratory of Particle and Radiation Imaging (Ministry of Education) and Department of Engineering Physics, Tsinghua University, Beijing 100084}
\author{L.~Wang}
\affiliation{Department of  Physics, Beijing Normal University, Beijing 100875}
\author{Q.~Wang}
\affiliation{Key Laboratory of Particle and Radiation Imaging (Ministry of Education) and Department of Engineering Physics, Tsinghua University, Beijing 100084}
\affiliation{Department of Physics, Tsinghua University, Beijing 100084}
\author{Y.~F.~Wang}
\affiliation{Key Laboratory of Particle and Radiation Imaging (Ministry of Education) and Department of Engineering Physics, Tsinghua University, Beijing 100084}
\author{Y.~X.~Wang}
\affiliation{School of Physics, Peking University, Beijing 100871}
\author{H.~T.~Wong}
\altaffiliation{Participating as a member of TEXONO Collaboration}
\affiliation{Institute of Physics, Academia Sinica, Taipei 11529}
\author{S.~Y.~Wu}
\affiliation{YaLong River Hydropower Development Company, Chengdu 610051}
\author{Y.~C.~Wu}
\affiliation{Key Laboratory of Particle and Radiation Imaging (Ministry of Education) and Department of Engineering Physics, Tsinghua University, Beijing 100084}
\author{H.~Y.~Xing}
\affiliation{College of Physics, Sichuan University, Chengdu 610065}
\author{R. Xu}
\affiliation{Key Laboratory of Particle and Radiation Imaging (Ministry of Education) and Department of Engineering Physics, Tsinghua University, Beijing 100084}
\author{Y.~Xu}
\affiliation{School of Physics, Nankai University, Tianjin 300071}
\author{T.~Xue}
\affiliation{Key Laboratory of Particle and Radiation Imaging (Ministry of Education) and Department of Engineering Physics, Tsinghua University, Beijing 100084}
\author{Y.~L.~Yan}
\affiliation{College of Physics, Sichuan University, Chengdu 610065}
\author{N.~Yi}
\affiliation{Key Laboratory of Particle and Radiation Imaging (Ministry of Education) and Department of Engineering Physics, Tsinghua University, Beijing 100084}
\author{C.~X.~Yu}
\affiliation{School of Physics, Nankai University, Tianjin 300071}
\author{H.~J.~Yu}
\affiliation{NUCTECH Company, Beijing 100084}
\author{J.~F.~Yue}
\affiliation{YaLong River Hydropower Development Company, Chengdu 610051}
\author{M.~Zeng}
\affiliation{Key Laboratory of Particle and Radiation Imaging (Ministry of Education) and Department of Engineering Physics, Tsinghua University, Beijing 100084}
\author{Z.~Zeng}
\affiliation{Key Laboratory of Particle and Radiation Imaging (Ministry of Education) and Department of Engineering Physics, Tsinghua University, Beijing 100084}
\author{B.~T.~Zhang}
\affiliation{Key Laboratory of Particle and Radiation Imaging (Ministry of Education) and Department of Engineering Physics, Tsinghua University, Beijing 100084}
\author{F.~S.~Zhang}
\affiliation{College of Nuclear Science and Technology, Beijing Normal University, Beijing 100875}
\author{L.~Zhang}
\affiliation{College of Physics, Sichuan University, Chengdu 610065}
\author{Z.~H.~Zhang}
\affiliation{Key Laboratory of Particle and Radiation Imaging (Ministry of Education) and Department of Engineering Physics, Tsinghua University, Beijing 100084}
\author{J.~Z.~Zhao}
\affiliation{Key Laboratory of Particle and Radiation Imaging (Ministry of Education) and Department of Engineering Physics, Tsinghua University, Beijing 100084}
\author{K.~K.~Zhao}
\affiliation{College of Physics, Sichuan University, Chengdu 610065}
\author{M.~G.~Zhao}
\affiliation{School of Physics, Nankai University, Tianjin 300071}
\author{J.~F.~Zhou}
\affiliation{YaLong River Hydropower Development Company, Chengdu 610051}
\author{Z.~Y.~Zhou}
\affiliation{Department of Nuclear Physics, China Institute of Atomic Energy, Beijing 102413}
\author{J.~J.~Zhu}
\affiliation{College of Physics, Sichuan University, Chengdu 610065}

\collaboration{CDEX Collaboration}
\noaffiliation

\date{\today}

\begin{abstract} Recently a dark matter--electron (DM--electron) paradigm has drawn much attention. Models beyond the standard halo model describing DM accelerated by high energy celestial bodies are under intense examination as well. In this Letter, a velocity components analysis (VCA) method dedicated to swift analysis of accelerated DM--electron interactions via semiconductor detectors is proposed and the first HPGe detector-based accelerated DM--electron analysis is realized. Utilizing the method, the first germanium based constraint on sub-GeV solar reflected DM--electron interaction is presented with the 205.4 kg$\cdot$day dataset from the CDEX-10 experiment. In the heavy mediator scenario, our result excels in the mass range of 5$-$15 keV/$c^2$, achieving a 3 orders of magnitude improvement comparing with previous semiconductor experiments. In the light mediator scenario, the strongest laboratory constraint for DM lighter than 0.1 MeV/$c^2$ is presented. The result proves the feasibility and demonstrates the vast potential of the VCA technique in future accelerated DM--electron analyses with semiconductor detectors.
\end{abstract}

\maketitle

\emph{Introduction.}— The enigma of dark matter (DM, denoted as $\chi$) remains a prevailing mystery in contemporary physics, potentially holding the key to understanding the nature and origin of the Universe~\cite{BERTONE2005279}. Previously, experiments probing DM within the mass range from GeV/$\rm{c}^2$ to TeV/$\rm{c}^2$ via DM--nucleus ($\chi$-$N$) scattering have been carried out extensively, such as XENON~\cite{xenonnt}, LUX-ZEPLIN~\cite{LZ}, PandaX~\cite{PandaX4T}, DarkSide~\cite{darkside}, SuperCDMS~\cite{cdmslite}, and CDEX~\cite{cdex0,cdex12014,cdex12016,cdex1,cdex1b2018,cdex10_tech,cdex102018,cdexdarkphoton,crdmcdex,c1b2019,cdexmidgal,cdex_exotic_dm,cdexchie}.  Recently, the DM--electron ($\chi$-$e$) scattering paradigm has drawn much attention. Comparing to nuclei, electrons can extract energy from light DM particles more efficiently, hence the probing ability improvement. Multiple experiments have conducted \chie analysis and pushed the \mchi reach down to $\sim$1 MeV/$c^2$~\cite{xenon10_100_chi_e,darkside_chi_e,damic_chi_e,damic_m_chi_e,sensei_chi_e,xenon1t_chi_e,pandaxii_chi_e,pandax_4T_chi_e,edelweiss_chi_e,cdmshvev_chi_e,cdexchie}. However, no trace of DM has been observed so far.

Previous \chie analyses were primarily dedicated to the DM described by the standard halo model (SHM)~\cite{SHM1,SHM2}. More recently, the significance of accelerated dark matter has been recognized. Prior to reaching the detector, DM particles may potentially interact with high-energy celestial bodies such as the Sun~\cite{AnHP_PRL2018,EmkenT_PRD2018,srdmanhp,srdmdamascus}, high-energy cosmic rays~\cite{crdmcdex,crdmphys,crdm_theory1,crdm_theory2,Bondarenko:2019vrb,crdm_theory4,HYPER_PRL2023}, blazars~\cite{BBDM_PRL,BBDM_SK}, supernovae~\cite{HuPK_PLB2017,LiJT_PRD2020,Monogem_PRD2023}, astrophysical neutrinos~\cite{vBDM,supernova_v,solar_neutrino,pbh_v}, atmospheric collisions~\cite{millicharged_PRD2020,atmosphere_PLB2022}, or black holes~\cite{PBHDM_chi-e,PBHDM_CDEX} and get accelerated, gaining sufficient energy to induce signals that surpass the detection threshold. This provides us with a good way to further enhance probing ability on \chie interactions. These accelerated DM models are collectively referred as accelerated DM.

For semiconductor detectors, calculations of accelerated DM--electron transition rates are considerably more complicated compared to noble gases with well tabulated wave functions~\cite{AtomWaveFunctions}, entailing more dedicated calculation techniques.

In this Letter, a novel method for the accelerated DM--electron interaction analysis on semiconductor detectors is proposed based on a modified version of the publicly available density functional theory (DFT) calculation package $\tt EXCEED$-$\tt DM$~\cite{ExceedDM020}. The package was originally aimed at the DM described by the SHM with Maxwell-Boltzmann distribution~\cite{am_theory}. In this work, we have modified it to analyze the detector response to DM with arbitrary analytical or numerical velocity distributions. Utilizing the approach, \chie constraints are derived for solar reflected dark matter (SRDM) with the 205.4 kg$\cdot$day dataset from high purity germanium (HPGe) detectors in the CDEX-10 experiment~\cite{cdexchie}.

\emph{Velocity components analysis method.}— Compared to noble gases, \chie calculations in semiconductors are considerably more complicated. By analyzing matrix elements depending solely on $\boldsymbol{q}$, and assuming electron energy levels to be spin independent, the \chie transition rate per target mass $R$ is determined with
\begin{align}
R&{=}\frac{2\pi\overline{\sigma}_{e}}{V\mu_{\chi e}^2m_{\chi}}\frac{\rho_{\chi}}{\rho_{T}}{\sum_{i,f}}{\int} \frac{{\rm d}^3q}{(2\pi)^3}{\left(\frac{f_e}{f_e^0}\right)^2}F^2_{\rm DM}g(\boldsymbol{q},\omega)|f_{i\rightarrow f}(\boldsymbol{q})|^2,\nonumber\\
\overline{\sigma}_{e}&{=}\frac{\mu_{\chi e}^2}{16\pi m_{\chi}^2m_e^2}\overline{|\mathcal{M}(q_0)|^2}{,}\ f_{i\rightarrow f}{=}{\int} {\rm d}^3xe^{i\boldsymbol{q}\cdot\boldsymbol{x}}\psi^\ast_f(\boldsymbol{x})\psi_i(\boldsymbol{x}),
\end{align}
where $\omega$ is the energy deposition, $\rho_{T}$ is the target density, $V$ is the target volume, $\rho_{\chi}$ is the local DM density which is taken to be 0.4 GeV/cm$^3$~\cite{ExceedDM_PRD}, $\mu_{\chi e}$ is the DM--electron reduced mass, $f_{i\rightarrow f}$ is the momentum transfer dependent crystal form factor, and $\overline{\sigma}_{e}$ is the reference cross section for free electron scattering~\cite{QEDark}. For simple DM models like the kinetically mixed dark photon or leptophilic scalar mediator models, the spin average matrix element squared $\overline{|\mathcal{M}(\boldsymbol{q})|^2}$ can be factorized as $\mathcal{M}(\boldsymbol{q})=\mathcal{M}(q_0)(f_e/f_e^0)F_{\rm DM}$, where the reference momentum transfer $q_0$ is taken to be $\alpha m_e$. $f_e/f_e^0$ is the screening factor discussed in detail in Ref.~\cite{ExceedDM_PRD}. $F_{\rm DM}$ is the dark matter form factor, where $F_{\rm DM}=1$ corresponds to pointlike interactions with heavy mediators or a magnetic dipole coupling, $F_{\rm DM}=q_0/q$ corresponds to an electric dipole coupling, and $F_{\rm DM}=(q_0/q)^2$ corresponds to massless or ultralight mediators. $g(\boldsymbol{q},\omega)=2\pi\int{\rm d}^3vf_{\chi}(\boldsymbol{v}_{\rm lab})\delta(\omega-\omega_q)$ is the kinematic factor~\cite{ExceedDM_PRD} that encapsulates the DM velocity distribution $f_{\chi}(\boldsymbol{v}_{\rm lab})$ in the lab frame. 

This calculation involves a six-dimensional integral, which is generally a numerically intensive task. For the commonly used Maxwell-Boltzmann (MB) distribution $f_{\chi}(\boldsymbol{v}_{\rm lab})=\frac{1}{N_0}e^{-\frac{|\boldsymbol{v}_{\rm lab}+\boldsymbol{v}_{\rm E}|^2}{v_0^2}}$, $g(\boldsymbol{q},\omega)$ is routinely evaluated analytically first to ease the computation. The kinematic factor for MB distribution in the SHM is determined with
\begin{equation}
\begin{aligned}
g(\boldsymbol{q},\omega)&=\frac{2\pi^2 v_0^2}{N_0}\frac{1}{q}(e^{-v_-^2/v_0^2}-e^{-v_{\rm esc}^2/v_0^2}),\\
v_-&={\rm min}\{\frac{1}{q}|\omega+\frac{q^2}{2m_\chi}+\boldsymbol{q}\cdot\boldsymbol{v}_{\rm E}|,v_{\rm esc}\},
\end{aligned}
\end{equation}
where the most probable velocity $v_0=220$ km/s, the Galactic escape velocity $v_{\rm esc}=544$ km/s, and the Earth's velocity in the Galactic rest frame $v_{\rm E}=232$ km/s~\cite{am_theory}.

However, a similar procedure is usually not applicable for accelerated DM models. Velocity distributions of accelerated DM models are obtained from Monte Carlo simulations or other methods, and in most cases, they cannot be integrated easily to obtain an analytical kinematic factor. Moreover, for accelerated DM models like SRDM, the velocity distributions depend not only on the DM masses but also on the cross sections. This means to perform a complete statistical analysis, we not only need to do the calculation numerically, but also have to perform the calculation repeatedly for each different set of DM masses and cross sections. This makes the analysis extremely time consuming. To bypass this barrier, a velocity components analysis (VCA) method is proposed.

\begin{figure}[!htbp]
\includegraphics[width=\linewidth]{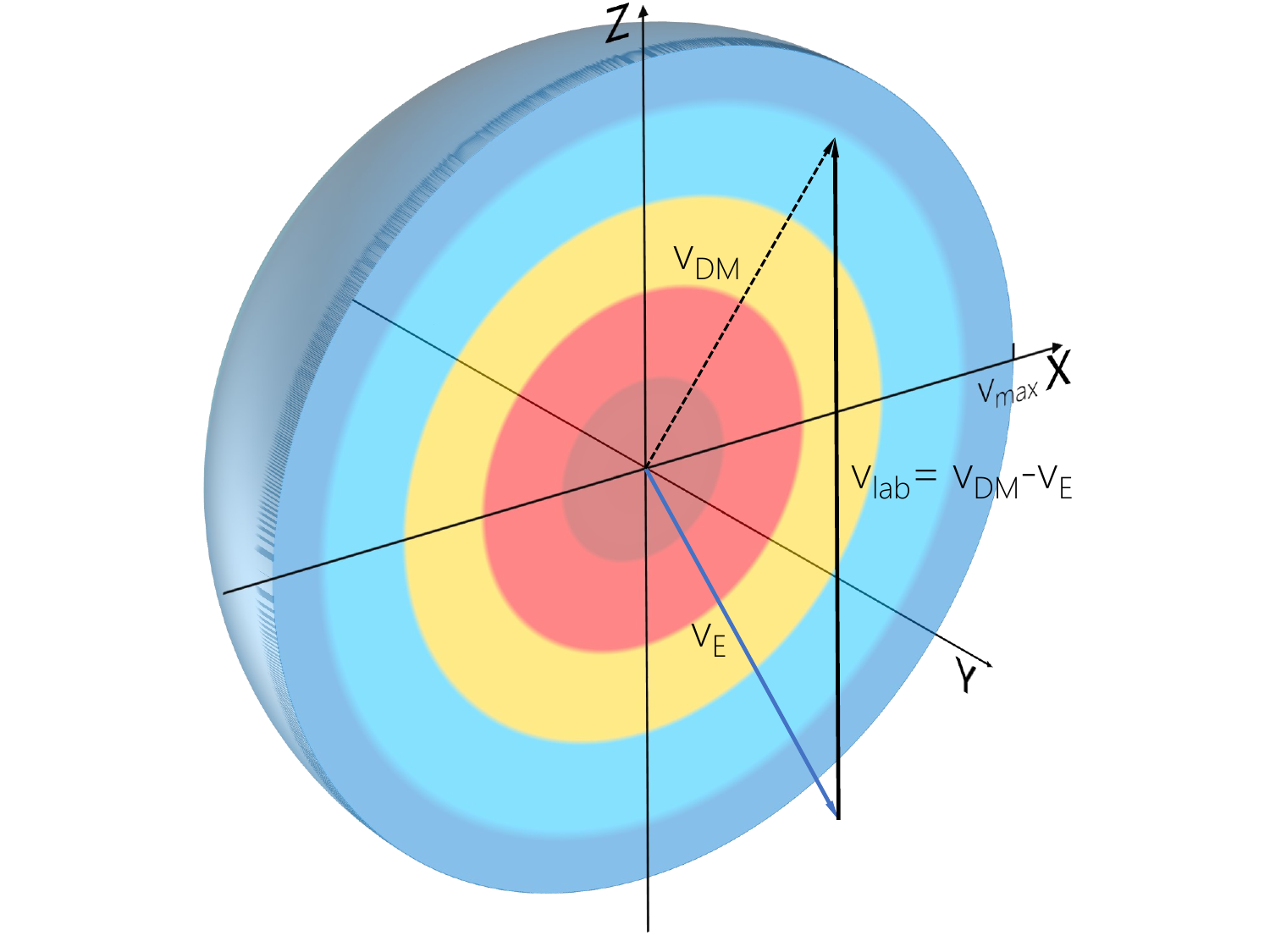}
\caption{
Velocity components of isotropic DM velocity distribution. $\boldsymbol v_{\rm DM}$ is DM particle's velocity. $\boldsymbol v_{\rm E}$ is the Earth's velocity relative to the DM model, and for SHM it's Earth's velocity with respect to the Galactic rest frame. $\boldsymbol v_{\rm lab}$ is the velocity seen by the lab. $\boldsymbol v_{\rm max}$ is the maximum velocity of the model. $\boldsymbol v_{\rm DM}=\boldsymbol v_{\rm lab}+\boldsymbol v_{\rm E}$.
}
\label{fig::Demo}
\end{figure}

As illustrated in Fig.~\ref{fig::Demo}, the sphere represents an arbitrary isotropic distribution of DM velocity $f'_{\chi}(\boldsymbol{v}_{\rm DM})$. $f_{\chi}(\boldsymbol{v}_{\rm lab})=f'_{\chi}(\boldsymbol{v}_{\rm lab}+\boldsymbol{v}_{\rm E})$ is the DM velocity distribution ``seen'' by the detectors in the lab frame moving relative to the model with a relative velocity $\boldsymbol v_{\rm E}$. Now if we ``peel'' the velocity distribution $f'_{\chi}(\boldsymbol{v}_{\rm DM})$ into $N$ layers of different velocity components in different velocity magnitude bins, as shown in Fig.~\ref{fig::Demo}, the detector response should be the summation of responses to all velocity magnitude bins. Then a binning distribution $h'_{k}(\boldsymbol{v}_{\rm DM})$ can be substituted for the component of $f'_{\chi}(\boldsymbol{v}_{\rm DM})$ in the $k_{\rm th}$ bin $f'_{k}(\boldsymbol{v}_{\rm DM})$ ($k\in[1,N]$). $h_{k}(\boldsymbol{v}_{\rm lab})=h'_{k}(\boldsymbol{v}_{\rm lab}+\boldsymbol{v}_{\rm E})$ and $f_{k}(\boldsymbol{v}_{\rm lab})=f'_{k}(\boldsymbol{v}_{\rm lab}+\boldsymbol{v}_{\rm E})$ are the boosted binning distribution and boosted $f'_{\chi}(\boldsymbol{v}_{\rm DM})$ component in the $k_{\rm th}$ bin. With a proper binning distribution $h'_{\chi}(\boldsymbol{v}_{\rm DM})$, the summation of responses to these bins should be close to the original total response.

The process can be expressed as follows:
\begin{equation}
\begin{aligned}
g(\boldsymbol q,\omega)&=2\pi\int{\rm d^3}v\sum_{k=1}^NA_kf_{k}(\boldsymbol{v}_{\rm lab})\delta(\omega-\omega_q)\\
&\approx2\pi\int{\rm d^3}v\sum_{k=1}^NA_kh_{k}(\boldsymbol{v}_{\rm lab})\delta(\omega-\omega_q)\\
&=\sum_{k=1}^NA_kg_k(\boldsymbol q,\omega),
\end{aligned}
\end{equation}
where $A_k=\int_{v_{{\rm min},k}}^{v_{{\rm max},k}}f'_{\chi}(\boldsymbol{v}_{\rm DM}){\rm d^3}v$ corresponds to the contribution from the $k_{\rm th}$ velocity bin with minimum and maximum speed of $v_{{\rm min},k}$ and $v_{{\rm max},k}$. $g_k(\boldsymbol q,\omega)$ is the kinematic factor of the $k_{\rm th}$ bin.

Given mathematically good binning distributions, the unweighted detector response to each layer can be calculated readily. With precalculated responses on hand, we only need to determine the contribution from each bin to reconstruct the total detector response for arbitrary velocity distributions:
\begin{align}
R&{=}\frac{\Phi_{\rm real}(m_{\chi},\overline\sigma_{e})}{\Phi_{\rm cal}(m_{\chi},\overline\sigma_{e})}\sum_{k=1}^NA_kR_{k},\\
R_k&{=}\frac{2\pi\overline{\sigma}_{e}}{V\mu_{\chi e}^2m_{\chi}}\frac{\rho_{\chi}}{\rho_{T}}{\sum_{i,f}}\int \frac{{\rm d}^3q}{(2\pi)^3}{\left(\frac{f_e}{f_e^0}\right)^2}F^2_{\rm DM}g_k(\boldsymbol{q},\omega)|f(\boldsymbol{q})|^2\nonumber,
\end{align}
where $\Phi_{\rm real}$ and $\Phi_{\rm cal}$ are the real flux and the flux of the reconstructed spectrum. 

The choice of the binning distribution is crucial. The binning distribution has to be analytical and easy to calculate to minimize the computation. The process may cause some deviations, but if the binning distribution is close to the original distribution, and the binning is sufficient, the deviations are anticipated to be acceptable. A reasonable choice is an inverse square distribution:
\begin{equation}
\begin{aligned}
h'_k(\boldsymbol{v}_{\rm DM})&{=}\frac{1}{C_k}\frac{1}{|\boldsymbol{v}_{\rm DM}|^2}\Theta((v_{{\rm max},k}{-}|\boldsymbol{v}_{\rm DM}|)(|\boldsymbol{v}_{\rm DM}|{-}v_{{\rm min},k})),\\
C_k&{=}4\pi({v_{{\rm max},k}}-{v_{{\rm min},k}}),
\end{aligned}
\end{equation}
where $\Theta(x)$ is the unit step function.

For inverse square distribution, the distribution shape is flat, which resembles most accelerated DM models, and contributions from different velocity magnitudes are uniform, which fits the feature of a binning method. The corresponding kinematic factor is
\begin{equation}
\begin{aligned}
g_k(\boldsymbol q,\omega)&=\frac{2\pi^2}{C_k q}{\rm ln}(\frac{{\rm max}\{{v_{{\rm max},k}}^2,d^2\}}{{\rm max}\{{v_{{\rm min},k}}^2,d^2\}}),\\
d&=\frac{1}{q}(\omega_q+\frac{q^2}{2m_\chi}+\boldsymbol{v}_{\rm E}\boldsymbol{q}).
\end{aligned}
\end{equation}

The $\tt EXCEED$-$\tt DM$ package~\cite{ExceedDM020} is modified to calculate contributions from different velocity magnitude bins. The package divides crystal electronic states into four categories: $core$, $valence$, $conduction$ and $free$ (denoted as c, v, cd, and f), and calculations of four transition types including v$\rightarrow$cd, v$\rightarrow$f, c$\rightarrow$cd, and c$\rightarrow$f are supported~\cite{ExceedDM_PRD}. Considering the maximum energy of the v$\rightarrow$cd process is less than 100 eV, which is lower than the typical Ge detector threshold, in this work only the v$\rightarrow$f, c$\rightarrow$cd, and c$\rightarrow$f processes are considered, and their contributions in different velocity magnitude bins are calculated separately.

\begin{figure}[!htbp]
\includegraphics[width=\linewidth]{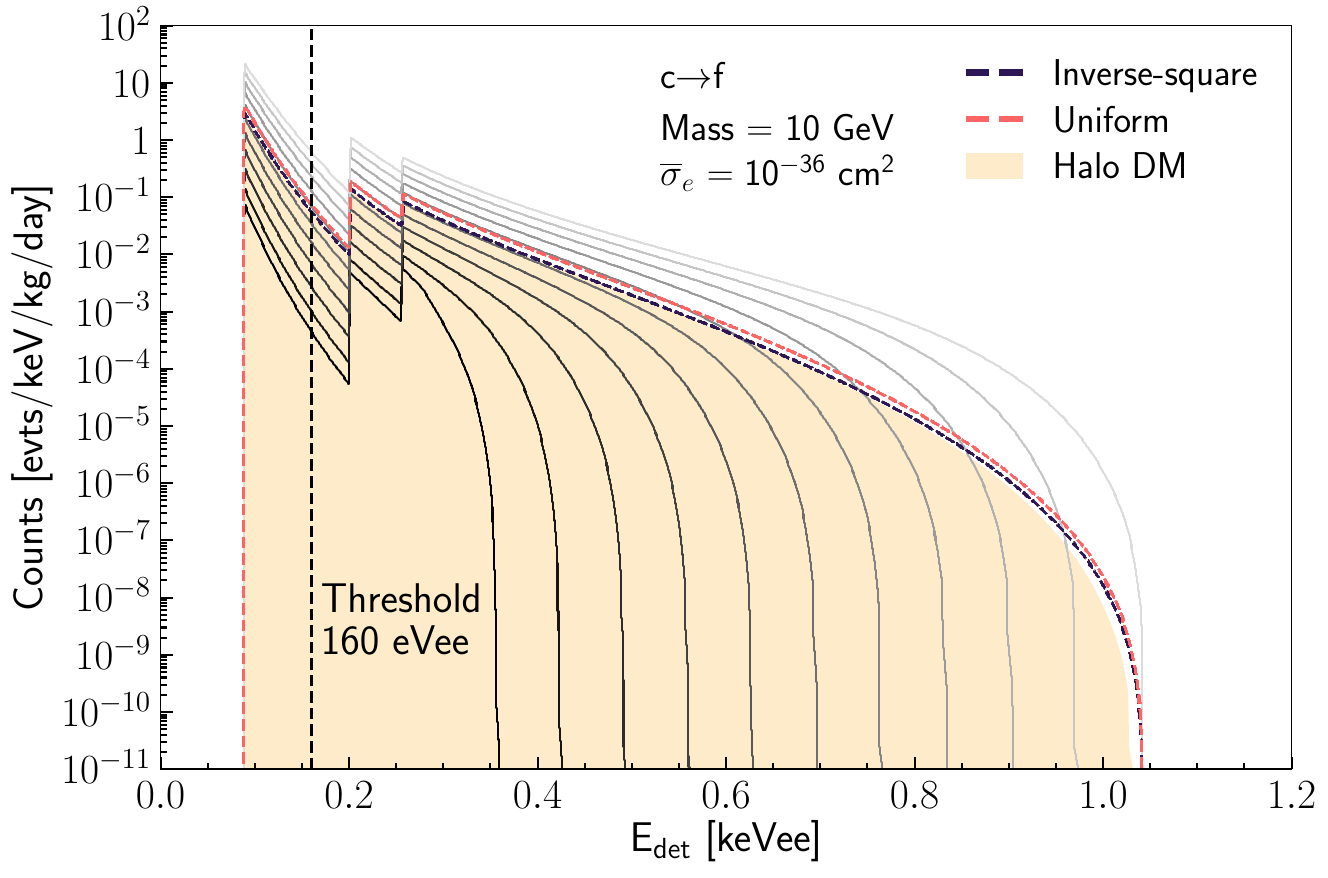}
\caption{
Reconstructed spectrum of the c$\rightarrow$f process for 10 GeV/$c^2$ DM from SHM in the heavy mediator scenario. The shaded area corresponds to the original result calculated by the $\tt EXCEED$-$\tt DM$ package~\cite{ExceedDM020}. The blue and red lines are reconstructed results using the inverse square and uniform distribution. Solid lines from darker to dimmer represent the contributions of 11 velocity bins with bin width of 50 km/s from 0$-$550 km/s calculated by the modified $\tt EXCEED$-$\tt DM$ package. The analysis threshold of CDEX-10 is represented by a black dashed line.
}
\label{fig::ReconstructedRes}
\end{figure}

Before applying the method to the accelerated DM analysis, it is first tested on the standard halo model. Contributions from eleven velocity magnitude bins with bin width of 50 km/s spanning from 0 km/s to 550 km/s are calculated separately and used for reconstruction. As depicted in Fig.~\ref{fig::ReconstructedRes}, the reconstructed spectrum accords well with the original spectrum. A uniform binning distribution $h'_k(\boldsymbol{v}_{\rm DM})= \rm {const}$ is tested as well. Compared to the uniform distribution, the inverse square distribution turns out to have a better accuracy, and the deviation of which near the threshold is $\sim$3\%. The following works are based on the inverse square binning distribution.

\begin{figure}[!htbp]
\includegraphics[width=\linewidth]{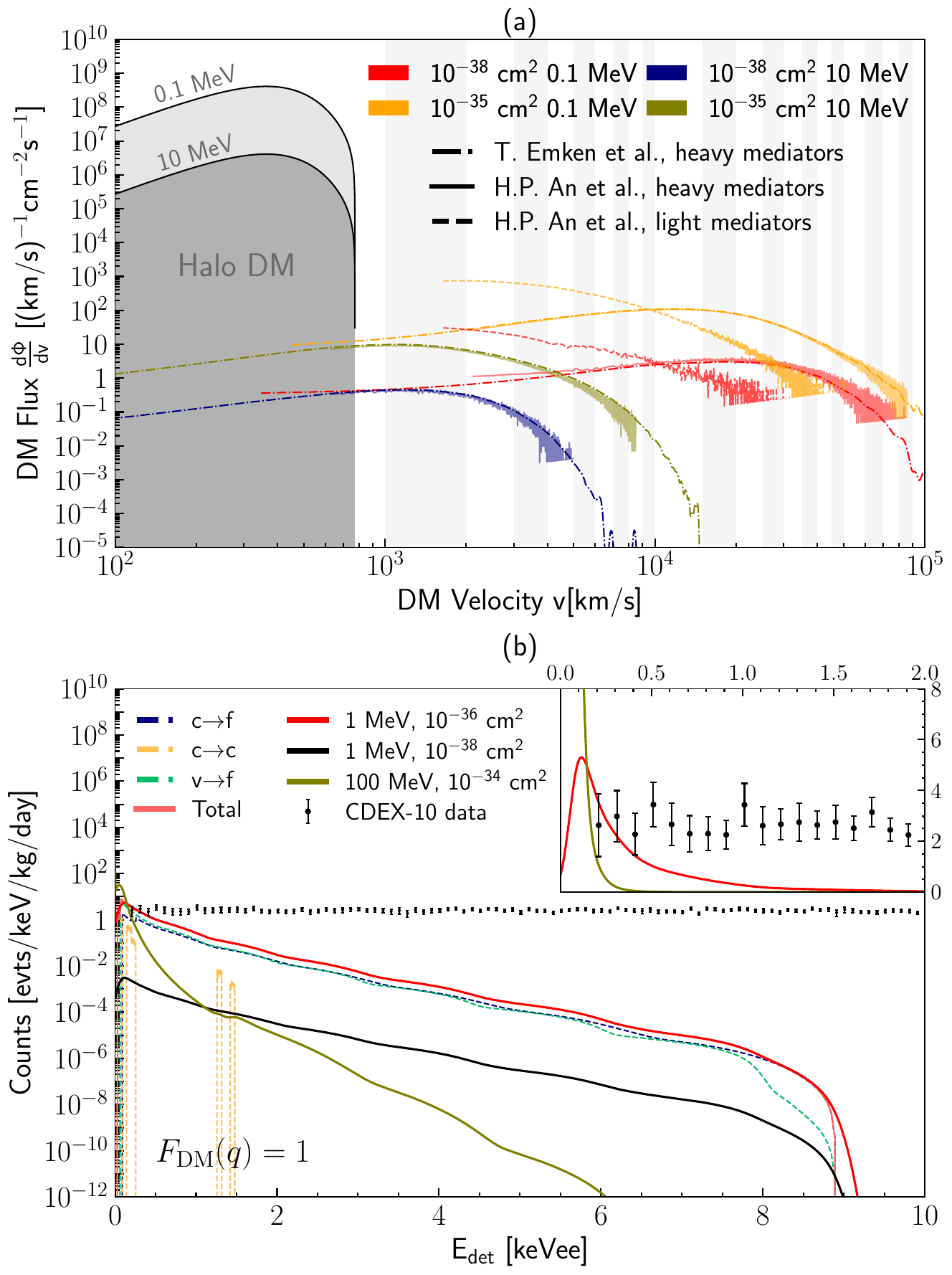}
\caption{
(a) SRDM flux distributions with different $m_{\chi}$ and $\overline{\sigma}_{e}$. SRDM flux in the light mediator scenario from Ref.~\cite{srdmanhp} is depicted in dashed lines. Other lines correspond to the heavy mediator scenario, and results from Ref.~\cite{srdmanhp} and Ref.~\cite{srdmdamascus} are consistent. The strips represent the binning of the velocity distributions. (b) Reconstructed HPGe detector response in the heavy mediator scenario to SRDM based on the distributions calculated by $\tt DaMaSCUS$-$\tt SUN$~\cite{damascuscode}. The detector's resolution is determined by $35.8 + 16.6\times\sqrt{E}\rm\ (eVee)$~\cite{crdmcdex}, where $E$ is in keVee. Experimental data from CDEX-10~\cite{crdmcdex} after efficiency correction with known radioactive peaks removed and zoomed details in 0.16$-$2.16 keVee are also depicted. The bin width is 100 eVee.
}
\label{fig::DaMaRes}
\end{figure}

\emph{Detector response to SRDM.}— Solar reflected DM can potentially be an important source of accelerated DM on Earth, and it might be a powerful instrument to enhance our DM probing ability~\cite{AnHP_PRL2018,EmkenT_PRD2018}. The velocity distributions of SRDM have already been thoroughly studied with different simulation approaches~\cite{srdmanhp,srdmdamascus}. In Ref.~\cite{srdmanhp}, the Sun is divided into 2,000 isotropic shells. The motion states of DM particles are updated each time they enter a new shell until they escape the Sun, and DM--electron interactions via both heavy and light mediators are studied. In Ref.~\cite{srdmdamascus}, the Monte Carlo package $\tt DaMaSCUS$-$\tt SUN$~\cite{damascuscode} is presented, in which the motions of DM particles in the Sun are directly calculated and updated according to the probability of contact interaction in current positions until they escape the Sun. As illustrated in Fig.~\ref{fig::DaMaRes}(a), in the heavy mediator scenario, distributions from both works are generally consistent. SRDM flux in the light mediator scenario from Ref.~\cite{srdmanhp} is also depicted. Before the analysis, the flux distributions have to be transformed to normalized velocity distributions: $f(v)=\frac{1}{N}\frac{{\rm d}\Phi}{v{\rm d}v}$, where $N$ is the normalization factor so that $\int f(v){\rm d}v=1$. Halo DM contributions are excluded from the SRDM distributions used in this work.

With the SRDM velocity distributions, the HPGe detector response to SRDM can be easily retrieved by performing the velocity components analysis. SRDM velocity distributions are first segmented as follows: ten 1,000 km/s bins in 0$-$$10^4$ km/s, eight 5,000 km/s bins in $10^4$$-$$5\times10^4$ km/s and five $10^4$ km/s bins in $5\times10^4$$-$$10^5$ km/s. Finer binning is adopted in the low velocity range because low velocity components account for the main part of SRDM distributions. Velocity components above $10^5$ km/s are conservatively ignored to avoid relativistic calculations. To verify if this binning method is sufficient, the spectrum of a test MB distribution with $v_0=3\times10^4$ km/s truncated at $10^5$ km/s is calculated directly, and reconstructed with current binning. The deviation of the reconstructed spectrum near the threshold is $\sim$0.2\% and negligible, confirming that current binning is sufficient. Finer binning will further reduce the deviation. Nevertheless, it's not necessary for our current data. Then contributions of v$\rightarrow$f, c$\rightarrow$cd, and c$\rightarrow$f process in these velocity bins are calculated separately and used in the reconstruction of the HPGe detector response to SRDM according to SRDM flux distributions. The reconstructed spectra of the v$\rightarrow$f, c$\rightarrow$cd, and c$\rightarrow$f processes, and the total spectra convolved with energy resolution are shown in Fig.~\ref{fig::DaMaRes}(b).

\emph{SRDM--electron analysis.}— With the reconstructed detector response to SRDM, and experimental data from CDEX-10, constraints on the SRDM--electron interaction can finally be established. The CDEX-10 experiment runs a 10-kg HPGe detector array in the China Jinping Underground Laboratory (CJPL) with a rock overburden of 2400 meters (6720 meters water equivalent)~\cite{cjpl,cjpl2}. Configuration of the experiment is described in detail in Refs.~\cite{cdex102018,cdex10_tech,cdexdarkphoton}. The analysis of the dataset follows the procedures established in our previous works~\cite{cdex1,cdex1b2018,cdex10_tech}, and the exposure of the dataset is 205.4 kg$\cdot$day~\cite{cdexdarkphoton,crdmcdex}. The energy calibration was performed with zero energy (defined by the random trigger events) and internal cosmogenic $K$-shell x-ray peaks at 8.98 keVee and 10.37 keVee from $^{65}$Zn and $^{68,71}$Ge. Physical events are identified with pedestal noise cut, physical event selection, and bulk or surface event discrimination~\cite{cdexbs}. Details of the procedures and efficiencies can be found in Refs.~\cite{cdex102018,cdex10_tech,cdexdarkphoton}. Finally the physical analysis threshold is set to be 160 eVee (``eVee'' represents the electron equivalent energy derived from energy calibration) where the combined signal efficiency, including the trigger efficiency and the efficiency for the pulse shape discrimination, is 4.5\%. The measured spectrum after efficiency correction and subtracting the contributions from $L$- and $M$-shell x-ray peaks derived from the corresponding $K$-shell line intensities~\cite{cdex102018,cdexdarkphoton,crdmcdex} is demonstrated in Fig.~\ref{fig::DaMaRes}(b). The background level of CDEX-10 achieves $\sim$2 counts keVee$^{-1}$kg$^{-1}$day$^{-1}$.

A minimum-$\chi^2$ analysis~\cite{cdex12014,cdexchie} is applied to the residual spectrum in the range of 0.16$-$12.06 keVee:
\begin{equation}
\begin{aligned}
\label{con:chi2}
\chi^2(m_{\chi},\overline{\sigma}_{e})=\sum_{i=1}^N\frac{[n_i-B-S_i(m_{\chi},\overline{\sigma}_{e})]^2}{\Delta_i^2},
\end{aligned}
\end{equation}
where $n_i$ and $\Delta_i$ are measured data and standard deviation with statistical and systematical components at the $i_{\rm th}$ energy bin, $S_i(m_{\chi},\overline{\sigma}_{e})$ is the predicted $\chi$-$e$ scattering rate, and $B$ is the assumed flat background from the Compton scattering of high energy gamma rays. The flat background assumption meets our understanding of the CDEX background model, and comparing with the background model with a slope, the deviation of the best-fit background is less than 3\% and negligible.

\begin{figure}[!htbp]
\includegraphics[width=\linewidth]{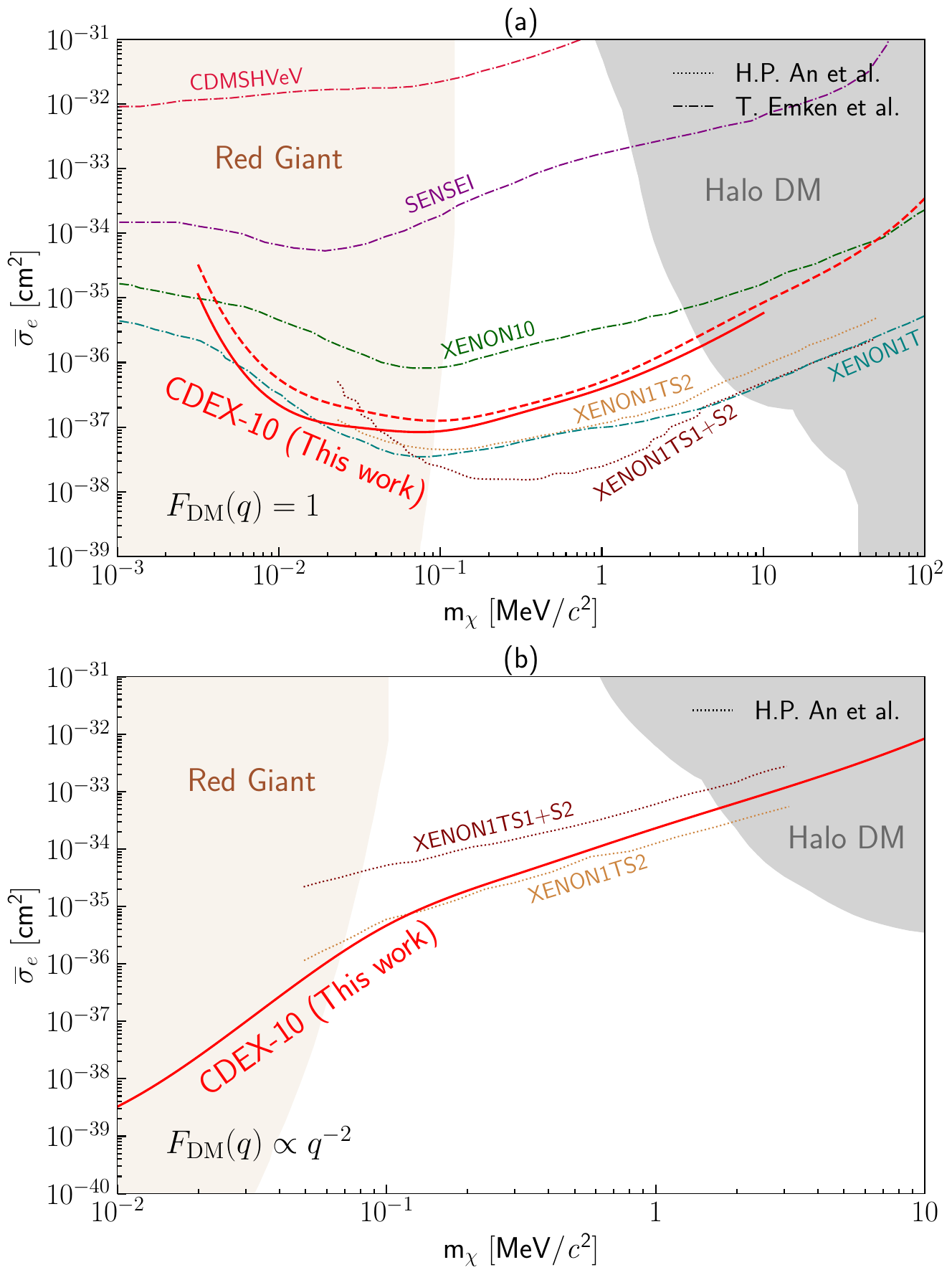}
\caption{
\chie constraints from the CDEX-10 experiment in the (a) heavy mediator  and (b) light mediator scenario. The red solid and dashed lines correspond to the CDEX-10 results derived with the SRDM flux from Ref.~\cite{srdmanhp} and Ref.~\cite{srdmdamascus}, respectively. SRDM constraints from XENON1TS2 and XENON1TS1+S2 presented by H.P. An $et\ al.$~\cite{srdmanhp} (dotted lines), and those from CDMSHVeV, SENSEI, XENON10, and XENON1T presented by T. Emken $et\ al.$~\cite{srdmdamascus} (dash-dotted lines), are also depicted. The gray shaded regions correspond to the current SHM constraints from direct search experiments~\cite{xenon1t_chi_e,pandaxii_chi_e,pandax_4T_chi_e,sensei_chi_e,damic_m_chi_e}. The brown shaded regions to the left are stellar cooling constraints from red giant (RG) stars for a dark photon-mediated model with $\alpha_D=0.5$ and $m_V = 3m_\chi$, where $m_V$ is the dark photon mass and $\alpha_D = e_D^2/4\pi$, where $e_D$ is the gauge coupling in the dark sector~\cite{redgiant}.}
\label{fig::Res}
\end{figure}

The 90$\%$ confidence level (C.L.) one-side upper limit exclusion lines of $\overline{\sigma}_{e}$ are derived~\cite{chisquare} using both velocity distributions presented by H.P. An $et\ al.$~\cite{srdmanhp} and T. Emken $et\ al.$~\cite{srdmdamascus}. We note that the Earth shielding effect is negligible at the level of our exclusion results~\cite{SiResponse}. Constraints from the CDEX-10 experiment and others presented by Ref.~\cite{srdmanhp} and Ref.~\cite{srdmdamascus} are depicted in Fig.~\ref{fig::Res}. The stellar cooling bounds from red giant (RG) stars for a dark photon-mediated model~\cite{redgiant} are superimposed. These are astrophysical constraints with model dependence. As shown in Fig.~\ref{fig::Res}(a) for the heavy mediator scenario, our limits are the most stringent in the mass range of 5$-$15 keV/$c^2$, and improve over previous semiconductor bounds by 3 orders of magnitude. As anticipated, constraints derived from both works~\cite{srdmanhp,srdmdamascus} accord with each other, and the deviation is within 30\% at a few MeV/$c^2$. In the light mediator scenario in Fig.~\ref{fig::Res}(b), our results provide the best laboratory constraint for DM lighter than 0.1 MeV/$c^2$, as well as the first semiconductor based SRDM result. The advances in sensitivities originate from the superior detector threshold and ultralow radiation environment.

For semiconductor results in Fig.~\ref{fig::Res}(a), the analysis is based on the method presented in Ref.~\cite{SiResponse} using $\tt QEDark$ package~\cite{QEDark}, which aims at the electronic states in valence and conduction bands (v$\rightarrow$cd). However, as depicted in Fig.~\ref{fig::DaMaRes}(b), for accelerated DM in high energy region (${>}100$ eV), the total spectrum is dominated by contributions from the previously ignored c$\rightarrow$f and v$\rightarrow$f process, which are no longer negligible as the maximum DM energy increases. Our result reveals that a more complete modeling of electronic states is necessary in the accelerated DM--electron analysis, especially for experiments with relatively high thresholds.

\emph{Summary.}— In this Letter, a velocity components analysis method to evaluate the detector response to dark matter particles with non-Maxwell-Boltzmann velocity distributions is proposed, and the first HPGe detector-based accelerated DM--electron analysis is realized. The method reflects a ``memory-for-time'' strategy: with a precalculated database, the detector response can be quickly reconstructed given a certain dark matter velocity distribution. The method is initially tested within the SHM, and then applied to the SRDM analysis.

Based on two different SRDM flux calculation approaches~\cite{srdmdamascus,srdmanhp}, and the data from the CDEX-10 experiment, we present leading laboratory constraints on SRDM--electron interactions in both heavy and light mediator scenario. This is also the first semiconductor based SRDM result in the light mediator scenario. The result reveals that complete modeling of electronic states is crucial in the accelerated DM--electron analysis, and demonstrates the feasibility and vast potential of the velocity components analysis method combined with a cutting-edge \chie calculation technique in future accelerated DM--electron analyses with semiconductor detectors.

This work opens a gateway for HPGe and other semiconductor detectors to perform a better analysis not only on SRDM, but also on other accelerated DM models~\cite{crdmphys,BBDM_SK,solar_neutrino,PBHDM_chi-e}. Our current research efforts target to upgrade this analysis method by adopting finer binning or taking a similar approach as $\tt QEDark$~\cite{QEDark} to save the crystal form factor as a function of $(\boldsymbol q,\omega)$ to further augment the calculation efficiency of VCA method. Studies of multiple accelerated DM models with HPGe detectors are currently being pursued.

This work was supported by the National Key Research and Development Program of China (Grants No. 2017YFA0402200 and No. 2022YFA1605000) and the National Natural Science Foundation of China (Grants No. 12322511, No. 12175112, No. 12005111, and No. 11725522). We acknowledge the Center of High performance computing, Tsinghua University for providing the facility support. We would like to thank CJPL and its staff for hosting and supporting the CDEX project. CJPL is jointly operated by Tsinghua University and Yalong River Hydropower Development Company.

\bibliography{SRDM.bib}

 \newcommand{\noop}[1]{}
\begin{thebibliography}{66}%
\makeatletter
\providecommand \@ifxundefined [1]{%
 \@ifx{#1\undefined}
}%
\providecommand \@ifnum [1]{%
 \ifnum #1\expandafter \@firstoftwo
 \else \expandafter \@secondoftwo
 \fi
}%
\providecommand \@ifx [1]{%
 \ifx #1\expandafter \@firstoftwo
 \else \expandafter \@secondoftwo
 \fi
}%
\providecommand \natexlab [1]{#1}%
\providecommand \enquote  [1]{``#1''}%
\providecommand \bibnamefont  [1]{#1}%
\providecommand \bibfnamefont [1]{#1}%
\providecommand \citenamefont [1]{#1}%
\providecommand \href@noop [0]{\@secondoftwo}%
\providecommand \href [0]{\begingroup \@sanitize@url \@href}%
\providecommand \@href[1]{\@@startlink{#1}\@@href}%
\providecommand \@@href[1]{\endgroup#1\@@endlink}%
\providecommand \@sanitize@url [0]{\catcode `\\12\catcode `\$12\catcode
  `\&12\catcode `\#12\catcode `\^12\catcode `\_12\catcode `\%12\relax}%
\providecommand \@@startlink[1]{}%
\providecommand \@@endlink[0]{}%
\providecommand \url  [0]{\begingroup\@sanitize@url \@url }%
\providecommand \@url [1]{\endgroup\@href {#1}{\urlprefix }}%
\providecommand \urlprefix  [0]{URL }%
\providecommand \Eprint [0]{\href }%
\providecommand \doibase [0]{http://dx.doi.org/}%
\providecommand \selectlanguage [0]{\@gobble}%
\providecommand \bibinfo  [0]{\@secondoftwo}%
\providecommand \bibfield  [0]{\@secondoftwo}%
\providecommand \translation [1]{[#1]}%
\providecommand \BibitemOpen [0]{}%
\providecommand \bibitemStop [0]{}%
\providecommand \bibitemNoStop [0]{.\EOS\space}%
\providecommand \EOS [0]{\spacefactor3000\relax}%
\providecommand \BibitemShut  [1]{\csname bibitem#1\endcsname}%
\let\auto@bib@innerbib\@empty
\bibitem [{\citenamefont {Bertone}\ \emph {et~al.}(2005)\citenamefont
  {Bertone}, \citenamefont {Hooper},\ and\ \citenamefont
  {Silk}}]{BERTONE2005279}%
  \BibitemOpen
  \bibfield  {author} {\bibinfo {author} {\bibfnamefont {G.}~\bibnamefont
  {Bertone}}, \bibinfo {author} {\bibfnamefont {D.}~\bibnamefont {Hooper}}, \
  and\ \bibinfo {author} {\bibfnamefont {J.}~\bibnamefont {Silk}},\ }\href
  {\doibase https://doi.org/10.1016/j.physrep.2004.08.031} {\bibfield
  {journal} {\bibinfo  {journal} {Phys. Rep.}\ }\textbf {\bibinfo {volume}
  {405}},\ \bibinfo {pages} {279} (\bibinfo {year} {2005})}\BibitemShut
  {NoStop}%
\bibitem [{\citenamefont {Aprile}\ \emph {et~al.}(2023)\citenamefont {Aprile}
  \emph {et~al.}}]{xenonnt}%
  \BibitemOpen
  \bibfield  {author} {\bibinfo {author} {\bibfnamefont {E.}~\bibnamefont
  {Aprile}} \emph {et~al.} (\bibinfo {collaboration} {XENON Collaboration}),\
  }\href {\doibase 10.1103/PhysRevLett.131.041003} {\bibfield  {journal}
  {\bibinfo  {journal} {Phys. Rev. Lett.}\ }\textbf {\bibinfo {volume} {131}},\
  \bibinfo {pages} {041003} (\bibinfo {year} {2023})}\BibitemShut {NoStop}%
\bibitem [{\citenamefont {Aalbers}\ \emph {et~al.}(2023)\citenamefont {Aalbers}
  \emph {et~al.}}]{LZ}%
  \BibitemOpen
  \bibfield  {author} {\bibinfo {author} {\bibfnamefont {J.}~\bibnamefont
  {Aalbers}} \emph {et~al.} (\bibinfo {collaboration} {LUX-ZEPLIN
  Collaboration}),\ }\href {\doibase 10.1103/PhysRevLett.131.041002} {\bibfield
   {journal} {\bibinfo  {journal} {Phys. Rev. Lett.}\ }\textbf {\bibinfo
  {volume} {131}},\ \bibinfo {pages} {041002} (\bibinfo {year}
  {2023})}\BibitemShut {NoStop}%
\bibitem [{\citenamefont {Meng}\ \emph {et~al.}(2021)\citenamefont {Meng} \emph
  {et~al.}}]{PandaX4T}%
  \BibitemOpen
  \bibfield  {author} {\bibinfo {author} {\bibfnamefont {Y.}~\bibnamefont
  {Meng}} \emph {et~al.} (\bibinfo {collaboration} {PandaX-4T Collaboration}),\
  }\href {\doibase 10.1103/PhysRevLett.127.261802} {\bibfield  {journal}
  {\bibinfo  {journal} {Phys. Rev. Lett.}\ }\textbf {\bibinfo {volume} {127}},\
  \bibinfo {pages} {261802} (\bibinfo {year} {2021})}\BibitemShut {NoStop}%
\bibitem [{\citenamefont {Agnes}\ \emph
  {et~al.}(2018{\natexlab{a}})\citenamefont {Agnes} \emph {et~al.}}]{darkside}%
  \BibitemOpen
  \bibfield  {author} {\bibinfo {author} {\bibfnamefont {P.}~\bibnamefont
  {Agnes}} \emph {et~al.} (\bibinfo {collaboration} {DarkSide Collaboration}),\
  }\href {\doibase 10.1103/PhysRevLett.121.081307} {\bibfield  {journal}
  {\bibinfo  {journal} {Phys. Rev. Lett.}\ }\textbf {\bibinfo {volume} {121}},\
  \bibinfo {pages} {081307} (\bibinfo {year} {2018}{\natexlab{a}})}\BibitemShut
  {NoStop}%
\bibitem [{\citenamefont {Agnese}\ \emph {et~al.}(2018)\citenamefont {Agnese}
  \emph {et~al.}}]{cdmslite}%
  \BibitemOpen
  \bibfield  {author} {\bibinfo {author} {\bibfnamefont {R.}~\bibnamefont
  {Agnese}} \emph {et~al.} (\bibinfo {collaboration} {SuperCDMS
  Collaboration}),\ }\href {\doibase 10.1103/PhysRevD.97.022002} {\bibfield
  {journal} {\bibinfo  {journal} {Phys. Rev. D}\ }\textbf {\bibinfo {volume}
  {97}},\ \bibinfo {pages} {022002} (\bibinfo {year} {2018})}\BibitemShut
  {NoStop}%
\bibitem [{\citenamefont {Liu}\ \emph {et~al.}(2014)\citenamefont {Liu} \emph
  {et~al.}}]{cdex0}%
  \BibitemOpen
  \bibfield  {author} {\bibinfo {author} {\bibfnamefont {S.~K.}\ \bibnamefont
  {Liu}} \emph {et~al.} (\bibinfo {collaboration} {CDEX Collaboration}),\
  }\href {\doibase 10.1103/PhysRevD.90.032003} {\bibfield  {journal} {\bibinfo
  {journal} {Phys. Rev. D}\ }\textbf {\bibinfo {volume} {90}},\ \bibinfo
  {pages} {032003} (\bibinfo {year} {2014})}\BibitemShut {NoStop}%
\bibitem [{\citenamefont {Yue}\ \emph {et~al.}(2014)\citenamefont {Yue} \emph
  {et~al.}}]{cdex12014}%
  \BibitemOpen
  \bibfield  {author} {\bibinfo {author} {\bibfnamefont {Q.}~\bibnamefont
  {Yue}} \emph {et~al.} (\bibinfo {collaboration} {CDEX Collaboration}),\
  }\href {\doibase 10.1103/PhysRevD.90.091701} {\bibfield  {journal} {\bibinfo
  {journal} {Phys. Rev. D}\ }\textbf {\bibinfo {volume} {90}},\ \bibinfo
  {pages} {091701} (\bibinfo {year} {2014})}\BibitemShut {NoStop}%
\bibitem [{\citenamefont {Zhao}\ \emph {et~al.}(2016)\citenamefont {Zhao} \emph
  {et~al.}}]{cdex12016}%
  \BibitemOpen
  \bibfield  {author} {\bibinfo {author} {\bibfnamefont {W.}~\bibnamefont
  {Zhao}} \emph {et~al.} (\bibinfo {collaboration} {CDEX Collaboration}),\
  }\href {\doibase 10.1103/PhysRevD.93.092003} {\bibfield  {journal} {\bibinfo
  {journal} {Phys. Rev. D}\ }\textbf {\bibinfo {volume} {93}},\ \bibinfo
  {pages} {092003} (\bibinfo {year} {2016})}\BibitemShut {NoStop}%
\bibitem [{\citenamefont {Zhao}\ \emph {et~al.}(2013)\citenamefont {Zhao} \emph
  {et~al.}}]{cdex1}%
  \BibitemOpen
  \bibfield  {author} {\bibinfo {author} {\bibfnamefont {W.}~\bibnamefont
  {Zhao}} \emph {et~al.} (\bibinfo {collaboration} {CDEX Collaboration}),\
  }\href {\doibase 10.1103/PhysRevD.88.052004} {\bibfield  {journal} {\bibinfo
  {journal} {Phys. Rev. D}\ }\textbf {\bibinfo {volume} {88}},\ \bibinfo
  {pages} {052004} (\bibinfo {year} {2013})}\BibitemShut {NoStop}%
\bibitem [{\citenamefont {Yang}\ \emph
  {et~al.}(2018{\natexlab{a}})\citenamefont {Yang} \emph
  {et~al.}}]{cdex1b2018}%
  \BibitemOpen
  \bibfield  {author} {\bibinfo {author} {\bibfnamefont {L.~T.}\ \bibnamefont
  {Yang}} \emph {et~al.} (\bibinfo {collaboration} {CDEX Collaboration}),\
  }\href {\doibase 10.1088/1674-1137/42/2/023002} {\bibfield  {journal}
  {\bibinfo  {journal} {Chin. Phys. C}\ }\textbf {\bibinfo {volume} {42}},\
  \bibinfo {eid} {023002} (\bibinfo {year} {2018}{\natexlab{a}})}\BibitemShut
  {NoStop}%
\bibitem [{\citenamefont {Jiang}\ \emph {et~al.}(2019)\citenamefont {Jiang}
  \emph {et~al.}}]{cdex10_tech}%
  \BibitemOpen
  \bibfield  {author} {\bibinfo {author} {\bibfnamefont {H.}~\bibnamefont
  {Jiang}} \emph {et~al.} (\bibinfo {collaboration} {CDEX Collaboration}),\
  }\href {\doibase 10.1007/s11433-018-8001-3} {\bibfield  {journal} {\bibinfo
  {journal} {Sci. China Phys. Mech. Astron.}\ }\textbf {\bibinfo {volume}
  {62}},\ \bibinfo {pages} {031012} (\bibinfo {year} {2019})}\BibitemShut
  {NoStop}%
\bibitem [{\citenamefont {Jiang}\ \emph {et~al.}(2018)\citenamefont {Jiang}
  \emph {et~al.}}]{cdex102018}%
  \BibitemOpen
  \bibfield  {author} {\bibinfo {author} {\bibfnamefont {H.}~\bibnamefont
  {Jiang}} \emph {et~al.} (\bibinfo {collaboration} {CDEX Collaboration}),\
  }\href {\doibase 10.1103/PhysRevLett.120.241301} {\bibfield  {journal}
  {\bibinfo  {journal} {Phys. Rev. Lett.}\ }\textbf {\bibinfo {volume} {120}},\
  \bibinfo {pages} {241301} (\bibinfo {year} {2018})}\BibitemShut {NoStop}%
\bibitem [{\citenamefont {She}\ \emph {et~al.}(2020)\citenamefont {She} \emph
  {et~al.}}]{cdexdarkphoton}%
  \BibitemOpen
  \bibfield  {author} {\bibinfo {author} {\bibfnamefont {Z.}~\bibnamefont
  {She}} \emph {et~al.} (\bibinfo {collaboration} {CDEX Collaboration}),\
  }\href {\doibase 10.1103/PhysRevLett.124.111301} {\bibfield  {journal}
  {\bibinfo  {journal} {Phys. Rev. Lett.}\ }\textbf {\bibinfo {volume} {124}},\
  \bibinfo {pages} {111301} (\bibinfo {year} {2020})}\BibitemShut {NoStop}%
\bibitem [{\citenamefont {Xu}\ \emph {et~al.}(2022)\citenamefont {Xu} \emph
  {et~al.}}]{crdmcdex}%
  \BibitemOpen
  \bibfield  {author} {\bibinfo {author} {\bibfnamefont {R.}~\bibnamefont {Xu}}
  \emph {et~al.} (\bibinfo {collaboration} {CDEX Collaboration}),\ }\href
  {\doibase 10.1103/PhysRevD.106.052008} {\bibfield  {journal} {\bibinfo
  {journal} {Phys. Rev. D}\ }\textbf {\bibinfo {volume} {106}},\ \bibinfo
  {pages} {052008} (\bibinfo {year} {2022})}\BibitemShut {NoStop}%
\bibitem [{\citenamefont {Yang}\ \emph {et~al.}(2019)\citenamefont {Yang} \emph
  {et~al.}}]{c1b2019}%
  \BibitemOpen
  \bibfield  {author} {\bibinfo {author} {\bibfnamefont {L.~T.}\ \bibnamefont
  {Yang}} \emph {et~al.} (\bibinfo {collaboration} {CDEX Collaboration}),\
  }\href {\doibase 10.1103/PhysRevLett.123.221301} {\bibfield  {journal}
  {\bibinfo  {journal} {Phys. Rev. Lett.}\ }\textbf {\bibinfo {volume} {123}},\
  \bibinfo {pages} {221301} (\bibinfo {year} {2019})}\BibitemShut {NoStop}%
\bibitem [{\citenamefont {Liu}\ \emph {et~al.}(2019)\citenamefont {Liu} \emph
  {et~al.}}]{cdexmidgal}%
  \BibitemOpen
  \bibfield  {author} {\bibinfo {author} {\bibfnamefont {Z.~Z.}\ \bibnamefont
  {Liu}} \emph {et~al.} (\bibinfo {collaboration} {CDEX Collaboration}),\
  }\href {\doibase 10.1103/PhysRevLett.123.161301} {\bibfield  {journal}
  {\bibinfo  {journal} {Phys. Rev. Lett.}\ }\textbf {\bibinfo {volume} {123}},\
  \bibinfo {pages} {161301} (\bibinfo {year} {2019})}\BibitemShut {NoStop}%
\bibitem [{\citenamefont {Dai}\ \emph {et~al.}(2022)\citenamefont {Dai} \emph
  {et~al.}}]{cdex_exotic_dm}%
  \BibitemOpen
  \bibfield  {author} {\bibinfo {author} {\bibfnamefont {W.~H.}\ \bibnamefont
  {Dai}} \emph {et~al.} (\bibinfo {collaboration} {CDEX Collaboration}),\
  }\href {\doibase 10.1103/PhysRevLett.129.221802} {\bibfield  {journal}
  {\bibinfo  {journal} {Phys. Rev. Lett.}\ }\textbf {\bibinfo {volume} {129}},\
  \bibinfo {pages} {221802} (\bibinfo {year} {2022})}\BibitemShut {NoStop}%
\bibitem [{\citenamefont {Zhang}\ \emph {et~al.}(2022)\citenamefont {Zhang}
  \emph {et~al.}}]{cdexchie}%
  \BibitemOpen
  \bibfield  {author} {\bibinfo {author} {\bibfnamefont {Z.~Y.}\ \bibnamefont
  {Zhang}} \emph {et~al.} (\bibinfo {collaboration} {CDEX Collaboration}),\
  }\href {\doibase 10.1103/PhysRevLett.129.221301} {\bibfield  {journal}
  {\bibinfo  {journal} {Phys. Rev. Lett.}\ }\textbf {\bibinfo {volume} {129}},\
  \bibinfo {pages} {221301} (\bibinfo {year} {2022})}\BibitemShut {NoStop}%
\bibitem [{\citenamefont {Essig}\ \emph {et~al.}(2017)\citenamefont {Essig}
  \emph {et~al.}}]{xenon10_100_chi_e}%
  \BibitemOpen
  \bibfield  {author} {\bibinfo {author} {\bibfnamefont {R.}~\bibnamefont
  {Essig}} \emph {et~al.},\ }\href {\doibase 10.1103/PhysRevD.96.043017}
  {\bibfield  {journal} {\bibinfo  {journal} {Phys. Rev. D}\ }\textbf {\bibinfo
  {volume} {96}},\ \bibinfo {pages} {043017} (\bibinfo {year}
  {2017})}\BibitemShut {NoStop}%
\bibitem [{\citenamefont {Agnes}\ \emph
  {et~al.}(2018{\natexlab{b}})\citenamefont {Agnes} \emph
  {et~al.}}]{darkside_chi_e}%
  \BibitemOpen
  \bibfield  {author} {\bibinfo {author} {\bibfnamefont {P.}~\bibnamefont
  {Agnes}} \emph {et~al.} (\bibinfo {collaboration} {DarkSide Collaboration}),\
  }\href {\doibase 10.1103/PhysRevLett.121.111303} {\bibfield  {journal}
  {\bibinfo  {journal} {Phys. Rev. Lett.}\ }\textbf {\bibinfo {volume} {121}},\
  \bibinfo {pages} {111303} (\bibinfo {year} {2018}{\natexlab{b}})}\BibitemShut
  {NoStop}%
\bibitem [{\citenamefont {Aguilar-Arevalo}\ \emph {et~al.}(2019)\citenamefont
  {Aguilar-Arevalo} \emph {et~al.}}]{damic_chi_e}%
  \BibitemOpen
  \bibfield  {author} {\bibinfo {author} {\bibfnamefont {A.}~\bibnamefont
  {Aguilar-Arevalo}} \emph {et~al.} (\bibinfo {collaboration} {DAMIC
  Collaboration}),\ }\href {\doibase 10.1103/PhysRevLett.123.181802} {\bibfield
   {journal} {\bibinfo  {journal} {Phys. Rev. Lett.}\ }\textbf {\bibinfo
  {volume} {123}},\ \bibinfo {pages} {181802} (\bibinfo {year}
  {2019})}\BibitemShut {NoStop}%
\bibitem [{\citenamefont {Arnquist}\ \emph {et~al.}(2023)\citenamefont
  {Arnquist} \emph {et~al.}}]{damic_m_chi_e}%
  \BibitemOpen
  \bibfield  {author} {\bibinfo {author} {\bibfnamefont {I.}~\bibnamefont
  {Arnquist}} \emph {et~al.} (\bibinfo {collaboration} {DAMIC-M
  Collaboration}),\ }\href {\doibase 10.1103/PhysRevLett.130.171003} {\bibfield
   {journal} {\bibinfo  {journal} {Phys. Rev. Lett.}\ }\textbf {\bibinfo
  {volume} {130}},\ \bibinfo {pages} {171003} (\bibinfo {year}
  {2023})}\BibitemShut {NoStop}%
\bibitem [{\citenamefont {Barak}\ \emph {et~al.}(2020)\citenamefont {Barak}
  \emph {et~al.}}]{sensei_chi_e}%
  \BibitemOpen
  \bibfield  {author} {\bibinfo {author} {\bibfnamefont {L.}~\bibnamefont
  {Barak}} \emph {et~al.} (\bibinfo {collaboration} {SENSEI Collaboration}),\
  }\href {\doibase 10.1103/PhysRevLett.125.171802} {\bibfield  {journal}
  {\bibinfo  {journal} {Phys. Rev. Lett.}\ }\textbf {\bibinfo {volume} {125}},\
  \bibinfo {pages} {171802} (\bibinfo {year} {2020})}\BibitemShut {NoStop}%
\bibitem [{\citenamefont {Aprile}\ \emph {et~al.}(2019)\citenamefont {Aprile}
  \emph {et~al.}}]{xenon1t_chi_e}%
  \BibitemOpen
  \bibfield  {author} {\bibinfo {author} {\bibfnamefont {E.}~\bibnamefont
  {Aprile}} \emph {et~al.} (\bibinfo {collaboration} {XENON Collaboration}),\
  }\href {\doibase 10.1103/PhysRevLett.123.251801} {\bibfield  {journal}
  {\bibinfo  {journal} {Phys. Rev. Lett.}\ }\textbf {\bibinfo {volume} {123}},\
  \bibinfo {pages} {251801} (\bibinfo {year} {2019})}\BibitemShut {NoStop}%
\bibitem [{\citenamefont {Cheng}\ \emph {et~al.}(2021)\citenamefont {Cheng}
  \emph {et~al.}}]{pandaxii_chi_e}%
  \BibitemOpen
  \bibfield  {author} {\bibinfo {author} {\bibfnamefont {C.}~\bibnamefont
  {Cheng}} \emph {et~al.} (\bibinfo {collaboration} {PandaX-II
  Collaboration}),\ }\href {\doibase 10.1103/PhysRevLett.126.211803} {\bibfield
   {journal} {\bibinfo  {journal} {Phys. Rev. Lett.}\ }\textbf {\bibinfo
  {volume} {126}},\ \bibinfo {pages} {211803} (\bibinfo {year}
  {2021})}\BibitemShut {NoStop}%
\bibitem [{\citenamefont {Li}\ \emph {et~al.}(2023)\citenamefont {Li} \emph
  {et~al.}}]{pandax_4T_chi_e}%
  \BibitemOpen
  \bibfield  {author} {\bibinfo {author} {\bibfnamefont {S.}~\bibnamefont {Li}}
  \emph {et~al.} (\bibinfo {collaboration} {PandaX Collaboration}),\ }\href
  {\doibase 10.1103/PhysRevLett.130.261001} {\bibfield  {journal} {\bibinfo
  {journal} {Phys. Rev. Lett.}\ }\textbf {\bibinfo {volume} {130}},\ \bibinfo
  {pages} {261001} (\bibinfo {year} {2023})}\BibitemShut {NoStop}%
\bibitem [{\citenamefont {Arnaud}\ \emph {et~al.}(2020)\citenamefont {Arnaud}
  \emph {et~al.}}]{edelweiss_chi_e}%
  \BibitemOpen
  \bibfield  {author} {\bibinfo {author} {\bibfnamefont {Q.}~\bibnamefont
  {Arnaud}} \emph {et~al.} (\bibinfo {collaboration} {EDELWEISS
  Collaboration}),\ }\href {\doibase 10.1103/PhysRevLett.125.141301} {\bibfield
   {journal} {\bibinfo  {journal} {Phys. Rev. Lett.}\ }\textbf {\bibinfo
  {volume} {125}},\ \bibinfo {pages} {141301} (\bibinfo {year}
  {2020})}\BibitemShut {NoStop}%
\bibitem [{\citenamefont {Amaral}\ \emph {et~al.}(2020)\citenamefont {Amaral}
  \emph {et~al.}}]{cdmshvev_chi_e}%
  \BibitemOpen
  \bibfield  {author} {\bibinfo {author} {\bibfnamefont {D.~W.}\ \bibnamefont
  {Amaral}} \emph {et~al.},\ }\href {\doibase 10.1103/PhysRevD.102.091101}
  {\bibfield  {journal} {\bibinfo  {journal} {Phys. Rev. D}\ }\textbf {\bibinfo
  {volume} {102}},\ \bibinfo {pages} {091101} (\bibinfo {year}
  {2020})}\BibitemShut {NoStop}%
\bibitem [{\citenamefont {Drukier}\ \emph {et~al.}(1986)\citenamefont
  {Drukier}, \citenamefont {Freese},\ and\ \citenamefont {Spergel}}]{SHM1}%
  \BibitemOpen
  \bibfield  {author} {\bibinfo {author} {\bibfnamefont {A.~K.}\ \bibnamefont
  {Drukier}}, \bibinfo {author} {\bibfnamefont {K.}~\bibnamefont {Freese}}, \
  and\ \bibinfo {author} {\bibfnamefont {D.~N.}\ \bibnamefont {Spergel}},\
  }\href {\doibase 10.1103/PhysRevD.33.3495} {\bibfield  {journal} {\bibinfo
  {journal} {Phys. Rev. D}\ }\textbf {\bibinfo {volume} {33}},\ \bibinfo
  {pages} {3495} (\bibinfo {year} {1986})}\BibitemShut {NoStop}%
\bibitem [{\citenamefont {Jungman}\ \emph {et~al.}(1996)\citenamefont
  {Jungman}, \citenamefont {Kamionkowski},\ and\ \citenamefont
  {Griest}}]{SHM2}%
  \BibitemOpen
  \bibfield  {author} {\bibinfo {author} {\bibfnamefont {G.}~\bibnamefont
  {Jungman}}, \bibinfo {author} {\bibfnamefont {M.}~\bibnamefont
  {Kamionkowski}}, \ and\ \bibinfo {author} {\bibfnamefont {K.}~\bibnamefont
  {Griest}},\ }\href {\doibase https://doi.org/10.1016/0370-1573(95)00058-5}
  {\bibfield  {journal} {\bibinfo  {journal} {Phys. Rep.}\ }\textbf {\bibinfo
  {volume} {267}},\ \bibinfo {pages} {195} (\bibinfo {year}
  {1996})}\BibitemShut {NoStop}%
\bibitem [{\citenamefont {An}\ \emph {et~al.}(2018)\citenamefont {An},
  \citenamefont {Pospelov}, \citenamefont {Pradler},\ and\ \citenamefont
  {Ritz}}]{AnHP_PRL2018}%
  \BibitemOpen
  \bibfield  {author} {\bibinfo {author} {\bibfnamefont {H.}~\bibnamefont
  {An}}, \bibinfo {author} {\bibfnamefont {M.}~\bibnamefont {Pospelov}},
  \bibinfo {author} {\bibfnamefont {J.}~\bibnamefont {Pradler}}, \ and\
  \bibinfo {author} {\bibfnamefont {A.}~\bibnamefont {Ritz}},\ }\href {\doibase
  10.1103/PhysRevLett.120.141801} {\bibfield  {journal} {\bibinfo  {journal}
  {Phys. Rev. Lett.}\ }\textbf {\bibinfo {volume} {120}},\ \bibinfo {pages}
  {141801} (\bibinfo {year} {2018})}\BibitemShut {NoStop}%
\bibitem [{\citenamefont {Emken}\ \emph {et~al.}(2018)\citenamefont {Emken},
  \citenamefont {Kouvaris},\ and\ \citenamefont {Nielsen}}]{EmkenT_PRD2018}%
  \BibitemOpen
  \bibfield  {author} {\bibinfo {author} {\bibfnamefont {T.}~\bibnamefont
  {Emken}}, \bibinfo {author} {\bibfnamefont {C.}~\bibnamefont {Kouvaris}}, \
  and\ \bibinfo {author} {\bibfnamefont {N.~G.}\ \bibnamefont {Nielsen}},\
  }\href {\doibase 10.1103/PhysRevD.97.063007} {\bibfield  {journal} {\bibinfo
  {journal} {Phys. Rev. D}\ }\textbf {\bibinfo {volume} {97}},\ \bibinfo
  {pages} {063007} (\bibinfo {year} {2018})}\BibitemShut {NoStop}%
\bibitem [{\citenamefont {An}\ \emph {et~al.}(2021)\citenamefont {An} \emph
  {et~al.}}]{srdmanhp}%
  \BibitemOpen
  \bibfield  {author} {\bibinfo {author} {\bibfnamefont {H.}~\bibnamefont {An}}
  \emph {et~al.},\ }\href {\doibase 10.1103/PhysRevD.104.103026} {\bibfield
  {journal} {\bibinfo  {journal} {Phys. Rev. D}\ }\textbf {\bibinfo {volume}
  {104}},\ \bibinfo {pages} {103026} (\bibinfo {year} {2021})},\ \bibinfo
  {note} {revised results in
  \href{https://doi.org/10.48550/arXiv.2108.10332}{arXiv:2108.10332}}\BibitemShut
  {NoStop}%
\bibitem [{\citenamefont {Emken}(2022)}]{srdmdamascus}%
  \BibitemOpen
  \bibfield  {author} {\bibinfo {author} {\bibfnamefont {T.}~\bibnamefont
  {Emken}},\ }\href {\doibase 10.1103/PhysRevD.105.063020} {\bibfield
  {journal} {\bibinfo  {journal} {Phys. Rev. D}\ }\textbf {\bibinfo {volume}
  {105}},\ \bibinfo {pages} {063020} (\bibinfo {year} {2022})}\BibitemShut
  {NoStop}%
\bibitem [{\citenamefont {Bringmann}\ \emph {et~al.}(2019)\citenamefont
  {Bringmann} \emph {et~al.}}]{crdmphys}%
  \BibitemOpen
  \bibfield  {author} {\bibinfo {author} {\bibfnamefont {T.}~\bibnamefont
  {Bringmann}} \emph {et~al.},\ }\href {\doibase
  10.1103/PhysRevLett.122.171801} {\bibfield  {journal} {\bibinfo  {journal}
  {Phys. Rev. Lett.}\ }\textbf {\bibinfo {volume} {122}},\ \bibinfo {pages}
  {171801} (\bibinfo {year} {2019})}\BibitemShut {NoStop}%
\bibitem [{\citenamefont {Dent}\ \emph {et~al.}(2020)\citenamefont {Dent},
  \citenamefont {Dutta}, \citenamefont {Newstead},\ and\ \citenamefont
  {Shoemaker}}]{crdm_theory1}%
  \BibitemOpen
  \bibfield  {author} {\bibinfo {author} {\bibfnamefont {J.~B.}\ \bibnamefont
  {Dent}}, \bibinfo {author} {\bibfnamefont {B.}~\bibnamefont {Dutta}},
  \bibinfo {author} {\bibfnamefont {J.~L.}\ \bibnamefont {Newstead}}, \ and\
  \bibinfo {author} {\bibfnamefont {I.~M.}\ \bibnamefont {Shoemaker}},\ }\href
  {\doibase 10.1103/PhysRevD.101.116007} {\bibfield  {journal} {\bibinfo
  {journal} {Phys. Rev. D}\ }\textbf {\bibinfo {volume} {101}},\ \bibinfo
  {pages} {116007} (\bibinfo {year} {2020})}\BibitemShut {NoStop}%
\bibitem [{\citenamefont {Cappiello}\ and\ \citenamefont
  {Beacom}(2019)}]{crdm_theory2}%
  \BibitemOpen
  \bibfield  {author} {\bibinfo {author} {\bibfnamefont {C.~V.}\ \bibnamefont
  {Cappiello}}\ and\ \bibinfo {author} {\bibfnamefont {J.~F.}\ \bibnamefont
  {Beacom}},\ }\href {\doibase 10.1103/PhysRevD.100.103011} {\bibfield
  {journal} {\bibinfo  {journal} {Phys. Rev. D}\ }\textbf {\bibinfo {volume}
  {100}},\ \bibinfo {pages} {103011} (\bibinfo {year} {2019})}\BibitemShut
  {NoStop}%
\bibitem [{\citenamefont {Bondarenko}\ \emph {et~al.}(2020)\citenamefont
  {Bondarenko}, \citenamefont {Boyarsky}, \citenamefont {Bringmann},
  \citenamefont {Hufnagel}, \citenamefont {Schmidt-Hoberg},\ and\ \citenamefont
  {Sokolenko}}]{Bondarenko:2019vrb}%
  \BibitemOpen
  \bibfield  {author} {\bibinfo {author} {\bibfnamefont {K.}~\bibnamefont
  {Bondarenko}}, \bibinfo {author} {\bibfnamefont {A.}~\bibnamefont
  {Boyarsky}}, \bibinfo {author} {\bibfnamefont {T.}~\bibnamefont {Bringmann}},
  \bibinfo {author} {\bibfnamefont {M.}~\bibnamefont {Hufnagel}}, \bibinfo
  {author} {\bibfnamefont {K.}~\bibnamefont {Schmidt-Hoberg}}, \ and\ \bibinfo
  {author} {\bibfnamefont {A.}~\bibnamefont {Sokolenko}},\ }\href {\doibase
  10.1007/JHEP03(2020)118} {\bibfield  {journal} {\bibinfo  {journal} {J. High
  Energy Phys.}\ }\textbf {\bibinfo {volume} {03}},\ \bibinfo {pages} {118}
  (\bibinfo {year} {2020})}\BibitemShut {NoStop}%
\bibitem [{\citenamefont {Guo}\ \emph {et~al.}(2020)\citenamefont {Guo},
  \citenamefont {Tsai}, \citenamefont {Wu},\ and\ \citenamefont
  {Yuan}}]{crdm_theory4}%
  \BibitemOpen
  \bibfield  {author} {\bibinfo {author} {\bibfnamefont {G.}~\bibnamefont
  {Guo}}, \bibinfo {author} {\bibfnamefont {Y.-L.~S.}\ \bibnamefont {Tsai}},
  \bibinfo {author} {\bibfnamefont {M.-R.}\ \bibnamefont {Wu}}, \ and\ \bibinfo
  {author} {\bibfnamefont {Q.}~\bibnamefont {Yuan}},\ }\href {\doibase
  10.1103/PhysRevD.102.103004} {\bibfield  {journal} {\bibinfo  {journal}
  {Phys. Rev. D}\ }\textbf {\bibinfo {volume} {102}},\ \bibinfo {pages}
  {103004} (\bibinfo {year} {2020})}\BibitemShut {NoStop}%
\bibitem [{\citenamefont {Elor}\ \emph {et~al.}(2023)\citenamefont {Elor},
  \citenamefont {McGehee},\ and\ \citenamefont {Pierce}}]{HYPER_PRL2023}%
  \BibitemOpen
  \bibfield  {author} {\bibinfo {author} {\bibfnamefont {G.}~\bibnamefont
  {Elor}}, \bibinfo {author} {\bibfnamefont {R.}~\bibnamefont {McGehee}}, \
  and\ \bibinfo {author} {\bibfnamefont {A.}~\bibnamefont {Pierce}},\ }\href
  {\doibase 10.1103/PhysRevLett.130.031803} {\bibfield  {journal} {\bibinfo
  {journal} {Phys. Rev. Lett.}\ }\textbf {\bibinfo {volume} {130}},\ \bibinfo
  {pages} {031803} (\bibinfo {year} {2023})}\BibitemShut {NoStop}%
\bibitem [{\citenamefont {Wang}\ \emph {et~al.}(2022)\citenamefont {Wang},
  \citenamefont {Granelli},\ and\ \citenamefont {Ullio}}]{BBDM_PRL}%
  \BibitemOpen
  \bibfield  {author} {\bibinfo {author} {\bibfnamefont {J.~W.}\ \bibnamefont
  {Wang}}, \bibinfo {author} {\bibfnamefont {A.}~\bibnamefont {Granelli}}, \
  and\ \bibinfo {author} {\bibfnamefont {P.}~\bibnamefont {Ullio}},\ }\href
  {\doibase 10.1103/PhysRevLett.128.221104} {\bibfield  {journal} {\bibinfo
  {journal} {Phys. Rev. Lett.}\ }\textbf {\bibinfo {volume} {128}},\ \bibinfo
  {pages} {221104} (\bibinfo {year} {2022})}\BibitemShut {NoStop}%
\bibitem [{\citenamefont {Granelli}\ \emph {et~al.}(2022)\citenamefont
  {Granelli} \emph {et~al.}}]{BBDM_SK}%
  \BibitemOpen
  \bibfield  {author} {\bibinfo {author} {\bibfnamefont {A.}~\bibnamefont
  {Granelli}} \emph {et~al.},\ }\href {\doibase 10.1088/1475-7516/2022/07/013}
  {\bibfield  {journal} {\bibinfo  {journal} {J. Cosmol. Astropart. Phys.}\
  }\textbf {\bibinfo {volume} {07}},\ \bibinfo {pages} {013} (\bibinfo {year}
  {2022})}\BibitemShut {NoStop}%
\bibitem [{\citenamefont {Hu}\ \emph {et~al.}(2017)\citenamefont {Hu},
  \citenamefont {Kusenko},\ and\ \citenamefont {Takhistov}}]{HuPK_PLB2017}%
  \BibitemOpen
  \bibfield  {author} {\bibinfo {author} {\bibfnamefont {P.-K.}\ \bibnamefont
  {Hu}}, \bibinfo {author} {\bibfnamefont {A.}~\bibnamefont {Kusenko}}, \ and\
  \bibinfo {author} {\bibfnamefont {V.}~\bibnamefont {Takhistov}},\ }\href
  {\doibase https://doi.org/10.1016/j.physletb.2017.02.035} {\bibfield
  {journal} {\bibinfo  {journal} {Phys. Lett. B}\ }\textbf {\bibinfo {volume}
  {768}},\ \bibinfo {pages} {18} (\bibinfo {year} {2017})}\BibitemShut
  {NoStop}%
\bibitem [{\citenamefont {Li}\ and\ \citenamefont {Lin}(2020)}]{LiJT_PRD2020}%
  \BibitemOpen
  \bibfield  {author} {\bibinfo {author} {\bibfnamefont {J.-T.}\ \bibnamefont
  {Li}}\ and\ \bibinfo {author} {\bibfnamefont {T.}~\bibnamefont {Lin}},\
  }\href {\doibase 10.1103/PhysRevD.101.103034} {\bibfield  {journal} {\bibinfo
   {journal} {Phys. Rev. D}\ }\textbf {\bibinfo {volume} {101}},\ \bibinfo
  {pages} {103034} (\bibinfo {year} {2020})}\BibitemShut {NoStop}%
\bibitem [{\citenamefont {Cappiello}\ \emph {et~al.}(2023)\citenamefont
  {Cappiello}, \citenamefont {Avis~Kozar},\ and\ \citenamefont
  {Vincent}}]{Monogem_PRD2023}%
  \BibitemOpen
  \bibfield  {author} {\bibinfo {author} {\bibfnamefont {C.~V.}\ \bibnamefont
  {Cappiello}}, \bibinfo {author} {\bibfnamefont {N.~P.}\ \bibnamefont
  {Avis~Kozar}}, \ and\ \bibinfo {author} {\bibfnamefont {A.~C.}\ \bibnamefont
  {Vincent}},\ }\href {\doibase 10.1103/PhysRevD.107.035003} {\bibfield
  {journal} {\bibinfo  {journal} {Phys. Rev. D}\ }\textbf {\bibinfo {volume}
  {107}},\ \bibinfo {pages} {035003} (\bibinfo {year} {2023})}\BibitemShut
  {NoStop}%
\bibitem [{\citenamefont {Jho}\ \emph {et~al.}(2021)\citenamefont {Jho} \emph
  {et~al.}}]{vBDM}%
  \BibitemOpen
  \bibfield  {author} {\bibinfo {author} {\bibfnamefont {Y.}~\bibnamefont
  {Jho}} \emph {et~al.},\ }\href@noop {} {\  (\bibinfo {year} {2021})},\
  \Eprint {http://arxiv.org/abs/2101.11262} {arXiv:2101.11262 [hep-ph]}
  \BibitemShut {NoStop}%
\bibitem [{\citenamefont {Das}\ and\ \citenamefont {Sen}(2021)}]{supernova_v}%
  \BibitemOpen
  \bibfield  {author} {\bibinfo {author} {\bibfnamefont {A.}~\bibnamefont
  {Das}}\ and\ \bibinfo {author} {\bibfnamefont {M.}~\bibnamefont {Sen}},\
  }\href {\doibase 10.1103/PhysRevD.104.075029} {\bibfield  {journal} {\bibinfo
   {journal} {Phys. Rev. D}\ }\textbf {\bibinfo {volume} {104}},\ \bibinfo
  {pages} {075029} (\bibinfo {year} {2021})}\BibitemShut {NoStop}%
\bibitem [{\citenamefont {Zhang}(2022)}]{solar_neutrino}%
  \BibitemOpen
  \bibfield  {author} {\bibinfo {author} {\bibfnamefont {Y.}~\bibnamefont
  {Zhang}},\ }\href {\doibase 10.1093/ptep/ptab156} {\bibfield  {journal}
  {\bibinfo  {journal} {Prog. Theor. Exp. Phys.}\ }\textbf {\bibinfo {volume}
  {2022}},\ \bibinfo {pages} {013B05} (\bibinfo {year} {2022})}\BibitemShut
  {NoStop}%
\bibitem [{\citenamefont {Chao}\ \emph {et~al.}(2021)\citenamefont {Chao},
  \citenamefont {Li},\ and\ \citenamefont {Liao}}]{pbh_v}%
  \BibitemOpen
  \bibfield  {author} {\bibinfo {author} {\bibfnamefont {W.}~\bibnamefont
  {Chao}}, \bibinfo {author} {\bibfnamefont {T.}~\bibnamefont {Li}}, \ and\
  \bibinfo {author} {\bibfnamefont {J.}~\bibnamefont {Liao}},\ }\href@noop {}
  {\  (\bibinfo {year} {2021})},\ \Eprint {http://arxiv.org/abs/2108.05608}
  {arXiv:2108.05608 [hep-ph]} \BibitemShut {NoStop}%
\bibitem [{\citenamefont {Plestid}\ \emph {et~al.}(2020)\citenamefont
  {Plestid}, \citenamefont {Takhistov}, \citenamefont {Tsai}, \citenamefont
  {Bringmann}, \citenamefont {Kusenko},\ and\ \citenamefont
  {Pospelov}}]{millicharged_PRD2020}%
  \BibitemOpen
  \bibfield  {author} {\bibinfo {author} {\bibfnamefont {R.}~\bibnamefont
  {Plestid}}, \bibinfo {author} {\bibfnamefont {V.}~\bibnamefont {Takhistov}},
  \bibinfo {author} {\bibfnamefont {Y.-D.}\ \bibnamefont {Tsai}}, \bibinfo
  {author} {\bibfnamefont {T.}~\bibnamefont {Bringmann}}, \bibinfo {author}
  {\bibfnamefont {A.}~\bibnamefont {Kusenko}}, \ and\ \bibinfo {author}
  {\bibfnamefont {M.}~\bibnamefont {Pospelov}},\ }\href {\doibase
  10.1103/PhysRevD.102.115032} {\bibfield  {journal} {\bibinfo  {journal}
  {Phys. Rev. D}\ }\textbf {\bibinfo {volume} {102}},\ \bibinfo {pages}
  {115032} (\bibinfo {year} {2020})}\BibitemShut {NoStop}%
\bibitem [{\citenamefont {Arguëlles}\ \emph {et~al.}(2022)\citenamefont
  {Arguëlles}, \citenamefont {Muñoz}, \citenamefont {Shoemaker},\ and\
  \citenamefont {Takhistov}}]{atmosphere_PLB2022}%
  \BibitemOpen
  \bibfield  {author} {\bibinfo {author} {\bibfnamefont {C.~A.}\ \bibnamefont
  {Arguëlles}}, \bibinfo {author} {\bibfnamefont {V.}~\bibnamefont {Muñoz}},
  \bibinfo {author} {\bibfnamefont {I.~M.}\ \bibnamefont {Shoemaker}}, \ and\
  \bibinfo {author} {\bibfnamefont {V.}~\bibnamefont {Takhistov}},\ }\href
  {\doibase https://doi.org/10.1016/j.physletb.2022.137363} {\bibfield
  {journal} {\bibinfo  {journal} {Phys. Lett. B}\ }\textbf {\bibinfo {volume}
  {833}},\ \bibinfo {pages} {137363} (\bibinfo {year} {2022})}\BibitemShut
  {NoStop}%
\bibitem [{\citenamefont {Calabrese}\ \emph {et~al.}(2022)\citenamefont
  {Calabrese} \emph {et~al.}}]{PBHDM_chi-e}%
  \BibitemOpen
  \bibfield  {author} {\bibinfo {author} {\bibfnamefont {R.}~\bibnamefont
  {Calabrese}} \emph {et~al.},\ }\href {\doibase 10.1103/PhysRevD.105.103024}
  {\bibfield  {journal} {\bibinfo  {journal} {Phys. Rev. D}\ }\textbf {\bibinfo
  {volume} {105}},\ \bibinfo {pages} {103024} (\bibinfo {year}
  {2022})}\BibitemShut {NoStop}%
\bibitem [{\citenamefont {Zhang}\ \emph {et~al.}(2023)\citenamefont {Zhang}
  \emph {et~al.}}]{PBHDM_CDEX}%
  \BibitemOpen
  \bibfield  {author} {\bibinfo {author} {\bibfnamefont {Z.~H.}\ \bibnamefont
  {Zhang}} \emph {et~al.} (\bibinfo {collaboration} {CDEX Collaboration}),\
  }\href {\doibase 10.1103/PhysRevD.108.052006} {\bibfield  {journal} {\bibinfo
   {journal} {Phys. Rev. D}\ }\textbf {\bibinfo {volume} {108}},\ \bibinfo
  {pages} {052006} (\bibinfo {year} {2023})}\BibitemShut {NoStop}%
\bibitem [{\citenamefont {Bunge}\ \emph {et~al.}(1993)\citenamefont {Bunge},
  \citenamefont {Barrientos},\ and\ \citenamefont {Bunge}}]{AtomWaveFunctions}%
  \BibitemOpen
  \bibfield  {author} {\bibinfo {author} {\bibfnamefont {C.}~\bibnamefont
  {Bunge}}, \bibinfo {author} {\bibfnamefont {J.}~\bibnamefont {Barrientos}}, \
  and\ \bibinfo {author} {\bibfnamefont {A.}~\bibnamefont {Bunge}},\ }\href
  {\doibase https://doi.org/10.1006/adnd.1993.1003} {\bibfield  {journal}
  {\bibinfo  {journal} {At. Data Nucl. Data Tables}\ }\textbf {\bibinfo
  {volume} {53}},\ \bibinfo {pages} {113} (\bibinfo {year} {1993})}\BibitemShut
  {NoStop}%
\bibitem [{\citenamefont {Trickle}\ and\ \citenamefont
  {kinzani}(2022)}]{ExceedDM020}%
  \BibitemOpen
  \bibfield  {author} {\bibinfo {author} {\bibfnamefont {T.}~\bibnamefont
  {Trickle}}\ and\ \bibinfo {author} {\bibnamefont {kinzani}},\ }\href
  {\doibase 10.5281/zenodo.7758287} {\enquote {\bibinfo {title}
  {{tanner-trickle/EXCEED-DM: EXCEED-DMv1.1.0}},}\ } (\bibinfo {year}
  {2022})\BibitemShut {NoStop}%
\bibitem [{\citenamefont {Smith}\ \emph {et~al.}(2007)\citenamefont {Smith}
  \emph {et~al.}}]{am_theory}%
  \BibitemOpen
  \bibfield  {author} {\bibinfo {author} {\bibfnamefont {M.~C.}\ \bibnamefont
  {Smith}} \emph {et~al.},\ }\href {\doibase 10.1111/j.1365-2966.2007.11964.x}
  {\bibfield  {journal} {\bibinfo  {journal} {Mon. Not. R. Astron. Soc.}\
  }\textbf {\bibinfo {volume} {379}},\ \bibinfo {pages} {755} (\bibinfo {year}
  {2007})}\BibitemShut {NoStop}%
\bibitem [{\citenamefont {Griffin}\ \emph {et~al.}(2021)\citenamefont {Griffin}
  \emph {et~al.}}]{ExceedDM_PRD}%
  \BibitemOpen
  \bibfield  {author} {\bibinfo {author} {\bibfnamefont {S.~M.}\ \bibnamefont
  {Griffin}} \emph {et~al.},\ }\href {\doibase 10.1103/PhysRevD.104.095015}
  {\bibfield  {journal} {\bibinfo  {journal} {Phys. Rev. D}\ }\textbf {\bibinfo
  {volume} {104}},\ \bibinfo {pages} {095015} (\bibinfo {year}
  {2021})}\BibitemShut {NoStop}%
\bibitem [{\citenamefont {Essig}\ \emph {et~al.}(2016)\citenamefont {Essig}
  \emph {et~al.}}]{QEDark}%
  \BibitemOpen
  \bibfield  {author} {\bibinfo {author} {\bibfnamefont {R.}~\bibnamefont
  {Essig}} \emph {et~al.},\ }\href {\doibase 10.1007/JHEP05(2016)046}
  {\bibfield  {journal} {\bibinfo  {journal} {J. High Energy Phys.}\ }\textbf
  {\bibinfo {volume} {05}},\ \bibinfo {pages} {046} (\bibinfo {year}
  {2016})}\BibitemShut {NoStop}%
\bibitem [{\citenamefont {Emken}(2021)}]{damascuscode}%
  \BibitemOpen
  \bibfield  {author} {\bibinfo {author} {\bibfnamefont {T.}~\bibnamefont
  {Emken}},\ }\href {\doibase DOI:10.5281/zenodo.5957388} {\enquote {\bibinfo
  {title} {{Dark Matter Simulation Code for Underground Scatterings - Sun
  Edition~(DaMaSCUS-SUN) [Code, v0.1.1]}},}\ }\bibinfo {howpublished}
  {Astrophysics Source Code Library record
  \href{https://ascl.net/2102.018}{[ascl:2102.018]}. The code can be found
  under \url{https://github.com/temken/damascus-sun}. Version 0.1.1 is archived
  as \href{https://doi.org/10.5281/zenodo.5957388}{DOI:10.5281/zenodo.5957388}}
  (\bibinfo {year} {2021})\BibitemShut {NoStop}%
\bibitem [{\citenamefont {Cheng}\ \emph {et~al.}(2017)\citenamefont {Cheng}
  \emph {et~al.}}]{cjpl}%
  \BibitemOpen
  \bibfield  {author} {\bibinfo {author} {\bibfnamefont {J.~P.}\ \bibnamefont
  {Cheng}} \emph {et~al.},\ }\href {\doibase
  10.1146/annurev-nucl-102115-044842} {\bibfield  {journal} {\bibinfo
  {journal} {Annu. Rev. Nucl. Part. Sci.}\ }\textbf {\bibinfo {volume} {67}},\
  \bibinfo {pages} {231} (\bibinfo {year} {2017})}\BibitemShut {NoStop}%
\bibitem [{\citenamefont {Wu}\ \emph {et~al.}(2013)\citenamefont {Wu} \emph
  {et~al.}}]{cjpl2}%
  \BibitemOpen
  \bibfield  {author} {\bibinfo {author} {\bibfnamefont {Y.~C.}\ \bibnamefont
  {Wu}} \emph {et~al.},\ }\href {\doibase 10.1088/1674-1137/37/8/086001}
  {\bibfield  {journal} {\bibinfo  {journal} {Chin. Phys. C}\ }\textbf
  {\bibinfo {volume} {37}},\ \bibinfo {pages} {086001} (\bibinfo {year}
  {2013})}\BibitemShut {NoStop}%
\bibitem [{\citenamefont {Yang}\ \emph
  {et~al.}(2018{\natexlab{b}})\citenamefont {Yang} \emph {et~al.}}]{cdexbs}%
  \BibitemOpen
  \bibfield  {author} {\bibinfo {author} {\bibfnamefont {L.~T.}\ \bibnamefont
  {Yang}} \emph {et~al.},\ }\href {\doibase 10.1016/j.nima.2017.12.078}
  {\bibfield  {journal} {\bibinfo  {journal} {Nucl. Instrum. Methods Phys.
  Res., Sect. A}\ }\textbf {\bibinfo {volume} {886}},\ \bibinfo {pages} {13}
  (\bibinfo {year} {2018}{\natexlab{b}})}\BibitemShut {NoStop}%
\bibitem [{\citenamefont {Feldman}\ and\ \citenamefont
  {Cousins}(1998)}]{chisquare}%
  \BibitemOpen
  \bibfield  {author} {\bibinfo {author} {\bibfnamefont {G.~J.}\ \bibnamefont
  {Feldman}}\ and\ \bibinfo {author} {\bibfnamefont {R.~D.}\ \bibnamefont
  {Cousins}},\ }\href {\doibase 10.1103/PhysRevD.57.3873} {\bibfield  {journal}
  {\bibinfo  {journal} {Phys. Rev. D}\ }\textbf {\bibinfo {volume} {57}},\
  \bibinfo {pages} {3873} (\bibinfo {year} {1998})}\BibitemShut {NoStop}%
\bibitem [{\citenamefont {Emken}\ \emph {et~al.}(2019)\citenamefont {Emken}
  \emph {et~al.}}]{SiResponse}%
  \BibitemOpen
  \bibfield  {author} {\bibinfo {author} {\bibfnamefont {T.}~\bibnamefont
  {Emken}} \emph {et~al.},\ }\href {\doibase 10.1088/1475-7516/2019/09/070}
  {\bibfield  {journal} {\bibinfo  {journal} {J. Cosmol. Astropart. Phys.}\
  }\textbf {\bibinfo {volume} {09}},\ \bibinfo {pages} {070} (\bibinfo {year}
  {2019})}\BibitemShut {NoStop}%
\bibitem [{\citenamefont {Chang}\ \emph {et~al.}(2021)\citenamefont {Chang},
  \citenamefont {Essig},\ and\ \citenamefont {Reinert}}]{redgiant}%
  \BibitemOpen
  \bibfield  {author} {\bibinfo {author} {\bibfnamefont {J.~H.}\ \bibnamefont
  {Chang}}, \bibinfo {author} {\bibfnamefont {R.}~\bibnamefont {Essig}}, \ and\
  \bibinfo {author} {\bibfnamefont {A.}~\bibnamefont {Reinert}},\ }\href
  {\doibase 10.1007/JHEP03(2021)141} {\bibfield  {journal} {\bibinfo  {journal}
  {J. High Energy Phys.}\ }\textbf {\bibinfo {volume} {03}},\ \bibinfo {pages}
  {141} (\bibinfo {year} {2021})}\BibitemShut {NoStop}%
\end{thebibliography}%

\end{document}